\documentclass[nofootinbib,showpacs,preprint,preprintnumbers,showkeys,amsmath,amssymb]{revtex4-1}
\pdfoutput=1
\usepackage[caption=false]{subfig}
\usepackage{amsmath}
\usepackage{amsfonts}
\usepackage{amssymb}
\usepackage{float}
\usepackage{color}
\usepackage{graphicx}
\usepackage{graphics}
\usepackage{hyperref}
\usepackage{adjustbox,array}
\usepackage{enumitem} 
\hypersetup{}
\usepackage[utf8x]{inputenc} 
\usepackage{epstopdf}
\usepackage{multirow}
\usepackage{slashed}
\usepackage{booktabs}
\usepackage{cancel}
\usepackage{mathrsfs}
\usepackage{csquotes}

\definecolor{rosso}{cmyk}{0,1,1,0.4}
\definecolor{rossos}{cmyk}{0,1,1,0.55}
\definecolor{rossoc}{cmyk}{0,1,1,0.2}
\definecolor{blu}{cmyk}{1,1,0,0.3}
\definecolor{blus}{cmyk}{1,1,0,0.6}
\definecolor{bluc}{cmyk}{1,1,0,0.1}
\definecolor{verde}{cmyk}{0.92,0,0.59,0.25}
\definecolor{verdec}{cmyk}{0.92,0,0.59,0.15}
\definecolor{verdes}{cmyk}{0.92,0,0.59,0.4}

\AtBeginDocument{
\heavyrulewidth=.08em
\lightrulewidth=.05em
\cmidrulewidth=.03em
\belowrulesep=.65ex
\belowbottomsep=0pt
\aboverulesep=.4ex
\abovetopsep=0pt
\cmidrulesep=\doublerulesep
\cmidrulekern=.5em
\defaultaddspace=.5em
}

\hypersetup{
 colorlinks=true,
 linkcolor=verdec,
 urlcolor=verdec,
 citecolor=verdec%gray
}

\newcommand{\eq}[1]{Eq.~(\ref{#1})}

\newcommand{\gsim}{\gtrsim}
\newcommand{\lsim}{\lesssim}

\newcommand{\lf}{\left(}
\newcommand{\ri}{\right)}

\newcommand{\nn}{\nonumber}

\renewcommand{\lg}{\mathscr{L}} % Amplitude
\newcommand{\mcl}{\mathcal{L}}
\newcommand{\mco}{\mathcal{O}}

\newcommand{\mcs}{\mathcal{S}}

\newcommand{\br}{{\rm{Br}}}

\newcommand{\jf}{ j_{\rm f}}
\newcommand{\eb}{ e^-_{\rm b}}

\newcommand{\tot}{{\rm tot}}
\newcommand{\lumtot}{\mathcal{L}_{\rm tot}}

\newcommand{\pb}{{\;{\rm pb}}}
\newcommand{\fb}{{\;{\rm fb}}}
\newcommand{\ab}{{\;{\rm ab}}}
\newcommand{\ifb}{{\,{\rm fb}^{-1}}}
\newcommand{\iab}{{\,{\rm ab}^{-1}}}

\newcommand{\gev}{{\,{\rm GeV}}}
\newcommand{\tev}{{\;{\rm TeV}}}

\newcommand{\beq}{\begin{equation}}
\newcommand{\eeq}{\end{equation}}
\newcommand{\bea}{\begin{eqnarray}}
\newcommand{\eea}{\end{eqnarray}}
\newcommand{\barr}{\begin{array}}
\newcommand{\earr}{\end{array}}
\newcommand{\bc}{\begin{center}}
\newcommand{\ec}{\end{center}}
\newcommand{\bit}{\begin{itemize}}
\newcommand{\eit}{\end{itemize}}
\newcommand{\ben}{\begin{enumerate}}
\newcommand{\een}{\end{enumerate}}
\newcommand{\al}{\alpha}
\newcommand{\bt}{\beta}

\newcommand{\Dt}{\Delta}

\newcommand{\sg}{\sigma}

\newcommand{\gm}{\gamma}
\newcommand{\Gm}{\Gamma}

\newcommand{\tauh}{{\tau_{\rm h}}}

\newcommand{\nsg}{N_{\rm s}}
\newcommand{\nbg}{N_{\rm b}}
\newcommand{\dbg}{\delta_{\rm bg}}
\newcommand{\xc}{x_{\rm cut}}

\newcommand{\wpm}{W^\pm}

\newcommand{\ee}      {{e^+ e^-}}

\newcommand{\ttau}      {{\tau^+\tau^-}} %{{\tau\tau}} %

\newcommand{\met}      {{E_T^{\rm miss}}}
\newcommand{\vmet}      {{\vec{E}_T^{\rm miss}}}

\newcommand{\fig}[1]{Fig.~\ref{#1}} % song comment

\begin{document}

\title{\color{verdes} Exploring lepton flavor violation phenomena 
of the $Z$ and Higgs bosons at electron-proton colliders}
\author{Adil Jueid}
\email{adiljueid@ibs.re.kr}
\address{Center for Theoretical Physics of the Universe, Institute for Basic Science (IBS), Daejeon, 34126, Republic of Korea}
\author{Jinheung Kim}
\email{jinheung.kim1216@gmail.com}
\address{Department of Physics, Konkuk University, Seoul 05029, Republic of Korea}
\author{Soojin Lee}
\email{soojinlee957@gmail.com}
\address{Department of Physics, Konkuk University, Seoul 05029, Republic of Korea}
\author{Jeonghyeon Song}
\email{jhsong@konkuk.ac.kr}
\address{Department of Physics, Konkuk University, Seoul 05029, Republic of Korea}
\author{Daohan Wang }
\email{wdh9508@gmail.com}
\address{Department of Physics, Konkuk University, Seoul 05029, Republic of Korea}

\begin{abstract}
We comprehensively study the potential for discovering lepton flavor violation (LFV) phenomena associated with the $Z$ and Higgs bosons at the LHeC and FCC-he. Our meticulous investigation reveals the remarkable suitability of electron-proton colliders for probing these rare new physics signals. This is due to the distinct advantages they offer, including negligible pileups, minimal QCD backgrounds, electron-beam polarization $P_e$, and the capability of distinguishing the charged-current from neutral-current processes.  In our pursuit of LFV of the $Z$ boson, we employ an innovative indirect probe, utilizing the $t$-channel mediation of the $Z$ boson in the process $p e^- \to j \tau^-$. For LFV in the Higgs sector, we scrutinize direct observations of the on-shell decays of $H\to e^+\tau^-$ and $H\to \mu^\pm\tau^\mp$ through the charged-current production of $H$. Focusing on $H\to e^+\tau^-$ proves highly efficient due to the absence of positron-related backgrounds in the charged-current modes at electron-proton colliders. Through a dedicated signal-to-background analysis with the boosted decision tree algorithm, we demonstrate that the LHeC with the total integrated luminosity of $1{\,{\rm ab}^{-1}}$ can put significantly lower $2\sigma$ bounds than the HL-LHC with $3{\,{\rm ab}^{-1}}$. Specifically, we find ${\rm{Br}}(Z\to e\tau)< 2.2 \times 10^{-7}$, ${\rm{Br}}(H\to e\tau) <1.7 \times 10^{-4} $, and ${\rm{Br}}(H\to \mu\tau) < 1.0 \times 10^{-4}$. Furthermore, our study uncovers the exceptional precision of the FCC-he in measuring the LFV signatures of the $Z$ and Higgs bosons, which indicates the potential for future discoveries in this captivating field.
\end{abstract}

\vspace{1cm}

\preprint{CTPU-PTC-23-18}

\maketitle
\tableofcontents

\section{Introduction}

Particle physics has made tremendous progress in the past few decades with the success of the Standard Model (SM) in explaining almost all experiments, including the discovery of the Higgs boson~\cite{ATLAS:2012yve,CMS:2012zhx}.
However, we are still on a quest for the ultimate theory of the universe, driven by two main reasons.
Firstly, the SM falls short in addressing fundamental questions that remain unanswered in our understanding of the universe. 
These include the nature of  dark matter~\cite{Navarro:1995iw,Bertone:2004pz},
the naturalness problem~\cite{Dimopoulos:1995mi,Chan:1997bi,Craig:2015pha}, 
the origin of neutrino masses, baryogenesis, and the metastability of the SM vacuum~\cite{Degrassi:2012ry}.
Secondly, tantalizing clues of physics beyond the SM (BSM) have surfaced.
Noteworthy examples include the muon anomalous magnetic moment~\cite{Muong-2:2006rrc,Muong-2:2021ojo,Aoyama:2019ryr},
the CDF $W$-boson mass~\cite{CDF:2022hxs},
the Cabibbo angle anomaly~\cite{Coutinho:2019aiy,Grossman:2019bzp,Belfatto:2019swo},
the excess in the diphoton mode around $96\gev$~\cite{CMS:2017yta}, and
the multi-lepton anomalies~\cite{vonBuddenbrock:2017gvy,Crivellin:2021ubm,Buddenbrock:2019tua}.
Any conclusive indication of BSM physics will bring new insights to our quest.

The study of lepton flavor violation (LFV) offers a remarkably clean pathway to explore BSM.
In the SM, LFV is highly suppressed  due to the Glashow-Iliopoulos-Maiani cancellation mechanism and the tiny neutrino masses~\cite{Glashow:1970gm}.
However, the observation of neutrino oscillations reveals that lepton flavor is not an exact symmetry,
making LFV an elusive yet captivating realm to investigate.
Numerous BSM models have been proposed to accommodate LFV,
such as massive neutrino models~\cite{Pilaftsis:1992st,Korner:1992an,Illana:2000ic,Arganda:2004bz,Arganda:2014dta,Arganda:2015naa,DeRomeri:2016gum,Arganda:2016zvc,Herrero:2018luu}, multi-Higgs doublet models~\cite{Bjorken:1977vt,Diaz-Cruz:1999sns,Iltan:2001au,Nomura:2020kcw},
supersymmetric models~\cite{Illana:2002tg,Arana-Catania:2013xma,Arana-Catania:2013xma,Arhrib:2012ax},
composite Higgs models~\cite{Agashe:2009di}, 
warped dimensional models~\cite{Perez:2008ee,Azatov:2009na,Albrecht:2009xr},
SMEFT~\cite{Calibbi:2021pyh},
and dark matter models~\cite{Jueid:2020yfj,Jueid:2023zxx}.
LFV signals can manifest at low energy scales
through the rare LFV decays of leptons and mesons, such as  $\mu \to e\gm$, $\mu\to eee$, $\tau\to e\gm /\mu \gm$,
$\mu \to e$ conversion in nuclei, and $\pi^0 \to \mu e$~\cite{Calibbi:2017uvl}.
Another intriguing category involves the LFV decays of the $Z$ or Higgs boson,
$Z\to L_\al L_\bt$ and $H \to L_\al L_\bt$ ($\al\neq\bt$) where 
$L_{\al,\bt} = e,\mu,\tau$.
In this work,
we focus on the latter,
as they present an enchanting research area within the context of high-energy colliders.

To provide a comprehensive overview of this intriguing topic, 
we present  in Table \ref{table:bounds} the current status and future prospects for the branching ratios of LFV decays of the $Z$ and Higgs boson
at the 95\% C.L. based on direct constraints. 
Notably, the $e\mu$ mode is stringently constrained by indirect measurements, such that 
$\br(Z\to e\mu)\lsim 10^{-12}$ from $\mu\to eee$~\cite{Dinh:2012bp,Calibbi:2021pyh} and
$\br(H\to e \mu)<\mco(10^{-8})$~\cite{Harnik:2012pb,Blankenburg:2012ex,Davidson:2012wn}
from $\mu\to e\gm$~\cite{MEG:2013oxv}.
However, the $e\tau$ and $\mu\tau$ modes still offer the possibility of sizable branching ratios,
indicating the potential for groundbreaking discoveries of LFV at high-energy colliders.

\begin{table}
%\begin{ruledtabular}
  {\renewcommand{\arraystretch}{1.2} \footnotesize
\begin{tabular}{|c|c|c||c|c|c|c|}
\hline
& current & future & &   current & \multicolumn{2}{c|}{ future}\\ \hline 
%Z1
\multirow{2}{*}{~$\br(Z\to e\mu)$~} & $<  2.62 \times 10^{-7}$ & $\lsim 10^{-8}$ 
& \multirow{2}{*}{~$\br(H\to e\mu)$~} & $<4.4 \times 10^{-5}$ & $<\mco(0.02)\%$ & $<1.2\times 10^{-5}$\\ 
& ATLAS~\cite{ATLAS:2022uhq}%139/fb 
&  ~FCC-ee-$Z$~\cite{Dam:2018rfz}~ &
&  CMS~\cite{CMS:2023pte}%138/fb
 &~HL-LHC~\cite{Banerjee:2016foh}~& $\ee$ collider~\cite{Qin:2017aju}\\ \hline
%Z2
\multirow{2}{*}{$\br(Z\to e\tau)$} & $<  5.0 \times 10^{-6}$ &  $\lsim 10^{-9}$ &   \multirow{2}{*}{$\br(H\to e\tau)$} & $<2.0\times 10^{-3}$ & $<\mco(0.5)\%$ & $<1.6\times 10^{-4}$  \\ 
& ATLAS~\cite{ATLAS:2021bdj}%139/fb
 & FCC-ee-$Z$~\cite{Dam:2018rfz}  & & ATLAS~\cite{ATLAS:2023mvd}%138/fb
  & HL-LHC~\cite{Banerjee:2016foh} & $\ee$ collider~\cite{Qin:2017aju} \\ \hline
%Z3
\multirow{2}{*}{$\br(Z\to \mu\tau)$} & $<   6.5 \times 10^{-6}$ & $\lsim 10^{-9}$
&   \multirow{2}{*}{$\br(H\to \mu\tau)$} & $<1.5\times 10^{-3}$ & $<1.0\times 10^{-3} $ & $\lsim 1.4\times 10^{-4}$  \\ 
& ATLAS~\cite{ATLAS:2021bdj}%139\fb
 & FCC-ee-$Z$~\cite{Dam:2018rfz} & 
&  CMS~\cite{CMS:2021rsq}%137/fb
 & HL-LHC~\cite{Barman:2022iwj} & $\ee$ collider~\cite{Qin:2017aju}  \\ \hline
\end{tabular}
}
%\end{ruledtabular}
\caption{\label{table:bounds}
The current status and future prospects of the direct bounds on the LFV decays of $Z$ and the Higgs boson at the 95\% C.L.
The prospects for the FCC-ee-$Z$ collider  are based on the configuration with $\sqrt{s}=88$-$95\gev$ and $\mcl=150\iab$,
while those for the $\ee$ collider are for $\sqrt{s}=240\gev$ and $\mcl = 5\iab$.}
%\end{ruledtabular}
\end{table}

Another important observation from Table \ref{table:bounds} is that
electron-proton colliders, such as the Large Hadron electron Collider (LHeC)~\cite{AbelleiraFernandez:2012cc,Bruening:2013bga,Agostini:2020fmq} and the Future Circular Collider electron-proton option (FCC-he)\cite{Abada:2019lih}, have not been  considered for LFV studies of the $Z$ and Higgs bosons.
The next-generation electron-proton colliders will play a crucial role in future high-energy hadron colliders, offering a unique advantage in providing high-precision data for accurately determining parton distribution functions of a proton.
In particular, the LHeC merits special attention 
due to its ability to operate simultaneously with the HL-LHC, 
made possible by the development of the energy recovery linac for the electron beam~\cite{Agostini:2020fmq}.

Unfortunately, the distinct advantages of electron-proton colliders in detecting rare BSM events have not been fully acknowledged in the existing literatures.
For instance, the electron beam mitigates issues related to pileup collisions.
This becomes crucial 
when probing rare BSM events,
which can be overshadowed  at the HL-LHC by the presence of over 150 pileup collisions per event.
The LHeC (FCC-he) offers a clean environment with approximately 0.1 (1) pileup collisions.
Additionally, electron-proton colliders provide a high level of control over backgrounds by disentangling the charged-current (CC) and neutral-current (NC) processes 
and enabling forward direction identification. 
Another advantage lies in the electron beam polarization $P_e$, which amplifies the cross sections in the CC channels.

These unique merits make electron-proton colliders exceptionally well-suited for investigating LFV signatures of the $Z$ and Higgs bosons.
Although directly measuring the decay of $Z\to e\tau/\mu\tau$ poses significant challenges due to their minuscule branching ratios below $\mco(10^{-6})$,
the electron beam provides an ingenious indirect probe 
through the process $p e^- \to j \tau^-$
mediated by the $Z$ boson in the $t$-channel diagram. 
To assess the sensitivity of this indirect channel, we will conduct a comprehensive detector simulation 
employing both a cut-based analysis and
a boosted decision tree (BDT) analysis~\cite{Roe:2004na}.

For the $H\to e\tau$ mode,
our focus will be on $H\to e^+\tau^-$ via the CC channel,
taking advantage of the absence of the positron-related backgrounds.
Encouragingly, the $H\to \mu^\pm\tau^\mp$ mode also benefits from a low background environment.
Given the extremely small production cross sections of the Higgs boson at the LHeC and FCC-he, 
 we will utilize a BDT analysis to enhance the signal-to-background discrimination 
and extract valuable information from the limited number of events. 
Our analysis will reveal that  the proposed channels at the LHeC with a total integrated luminosity of $\lumtot=1\iab$ 
yield significantly lower bounds 
on the branching ratios than the projected bounds at the HL-LHC with $\lumtot=3\iab$. 
Furthermore, the FCC-he exhibits even greater potential in this regard.
These findings represent a novel and significant contribution to the topic,
highlighting the untapped potential of electron-proton colliders in probing the LFV phenomena.

The paper is structured as follows:
In Sec.~\ref{sec:Formalism}, we provide a model-independent formalism 
for the LFV phenomena of the $Z$ and Higgs bosons.
After calculating the production cross sections of $H$, $Z$, $\wpm$, and $jj\nu$ in the SM, 
we discuss all the possible channels to probe the LFV signatures 
at electron-proton colliders.
Based on this, we propose the most promising channels.
In Sec.~\ref{sec:analysis}, we present our signal-to-background analysis using a full detector simulation.
Finally, we summarize our findings and conclusions in Sec.~\ref{sec:Conclusions}.

%%%%
\section{Formalism}
\label{sec:Formalism}
%%%%

In our exploration of LFV signatures associated with the $Z$ and Higgs bosons at electron-proton colliders, 
we employ a model-independent approach. 
To characterize LFV in the $Z$ boson sector, we consider the following LFV couplings involving the $Z$ boson, an electron, and a tau lepton:
\begin{equation}
    \label{eq:lg:Z}
    - \lg_{\rm LFV}^Z = Z_\mu \left[  \bar{\tau} \gm^\mu \left(
    C_{ \tau e}^L P_L + C_{\tau e}^R P_R 
    \right)  e
    +
    \bar{e} \gm^\mu \left(
    C_{e \tau}^L P_L + C_{e \tau}^R P_R 
    \right)  \tau
     \right]   ,
\end{equation}
where $P_{R,L} = (1\pm \gm^5)/2$. 
The couplings $C_{e\tau,\tau e}^{L,R}$ are connected to the LFV branching ratio according to
\bea
\left| C_{ \tau e}^L \right|^2 + \left| C_{ \tau e}^R \right|^2  
+
\left| C_{e \tau}^L \right|^2 + \left| C_{ e\tau}^R \right|^2
= \frac{24\pi \Gm^\tot_Z}{m_Z} \,\br(Z \to e \tau).
\eea
In this study, our focus lies on the indirect process utilizing the $t$-channel mediation of the $Z$ boson in the process $p e^- \to j \tau^-$,
which examines the couplings $C_{\tau e}^{L/R}$.
However, it is necessary to reinterpret the current limit of $\br(Z \to e\tau)<5.0 \times 10^{-6}$ in terms of  $C_{\tau e}^{L,R}$.
We take a reasonable assumption of $\big| C_{ \tau e}^{L/R} \big| =\big| C_{  e\tau}^{L/R}\big| $
and obtain the constraint of
\bea
\label{eq:CZ:bound}
\sqrt{|C_{\tau e}^R|^2+ |C_{\tau e}^L|^2}< 2.27 \times 10^{-3}.
\eea

Regarding the LFV Yukawa couplings of the Higgs boson, we introduce the following interaction Lagrangian:
\begin{equation}
\label{eq:lg:H}
-    \lg_{\rm LFV}^H =  Y_{\al\bt}\ \bar{L}_\al P_R L_\bt H + {\rm H.c.} , \quad (\al \neq \bt),
\end{equation}
where $L_{\al,\bt}=e,\mu,\tau$.
The relationship between $Y_{\al\bt}$ and $\br(H \to L_\al L_\bt)$ is given by
\bea
\left| Y_{\al\bt} \right|^2 + \left| Y_{\bt\al} \right|^2 = \frac{8 \pi }{m_H} \, \frac{\br(H \to L_\al L_\bt)}{1-\br(H \to L_\al L_\bt)}\,
\Gm_H^\tot.
\eea
The current constraints of $\br(H \to e\tau)< 2.2 \times 10^{-3}$ and $\br(H \to \mu\tau)< 1.5 \times 10^{-3}$ imply
\begin{align}
\sqrt{\left| Y_{ e\tau} \right|^2 + \left| Y_{ \tau e} \right|^2 } &< 1.30 \times 10^{-3},
\\[3pt] \nn
\sqrt{\left| Y_{ \mu\tau} \right|^2 + \left| Y_{\tau \mu} \right|^2 } &< 1.13 \times 10^{-3}.
\end{align}

\begin{table}[!t]
\begin{center}
\begin{tabular}{|c||c|c|c|c|c|c|}
\toprule
 \multicolumn{7}{|c|}{~~~Total SM cross sections in pb at the LHeC with $E_e=50\gev$ and $E_p=7\tev$}\\
\toprule
 & & Higgs & \multicolumn{2}{c|}{$\wpm$} & $Z$ & multijets \\ \hline
\multirow{2}{*}{~~~CC~~~} & & ~$\sg(H +\jf\nu_e)$~ & ~$\sg(W^- +\jf\nu_e )$~ & ~$\sg(W^+ + \jf\nu_e )$~ & ~$\sg(Z +\jf\nu_e )$~~~ 
 & ~~~$\sg(jj\nu_e)$~~~ \\ \cline{2-7}
& ~~$P_e=0$~~ & $0.081$ & $0.925$ & 0 & $0.456$ &  $72.53$ \\
& ~~$P_e=-80\%$~~ & $0.145$ & $1.657$ & 0 & $0.824$ &  $130.9$ \\ \hline
\multirow{2}{*}{NC} & & $\sg(H +\jf\eb)$ & $\sg(W^-+\jf \eb )$ &$\sg(W^++\jf \eb)$ & $\sg(Z+\jf \eb)$ 
 & $\sg(jj \eb)$ \\ \cline{2-7}
& $P_e=0$ & $0.0144$ & $1.031$ & $1.112$ & $0.244$  & $961.0$ \\
& ~~$P_e=-80\%$~~ & $0.0171$ & $1.433$ & $1.478$ & $0.3138$ &  $1019$ \\ \hline
\toprule
 \multicolumn{7}{|c|}{Total SM cross sections in pb at the FCC-he with $E_e=60\gev$ and $E_p=50\tev$}\\
\toprule
\multirow{2}{*}{~~~CC~~~} & &$\sg(H+\jf \nu_e)$ & $\sg(W^- +\jf\nu_e)$~~ & ~~$\sg(W^+ +\jf\nu_e )$~~ & ~~$\sg(Z+\jf \nu_e)$~~ 
 & ~~$\sg(jj\nu_e)$~~ \\ \cline{2-7}
& $P_e=0$ & $0.335$ & $4.253$ & 0 & $2.205 $ &  $235.2 $ \\ 
& $P_e=-80\%$ & $0.604$ & $7.618 $ & 0 & $3.969 $ &  $424 $ \\ \hline
\multirow{2}{*}{NC} & & $\sg(H +\jf\eb)$ & $\sg(W^-+\jf \eb )$ &$\sg(W^+ +\jf\eb )$ & $\sg(Z+ \jf\eb)$ & $\sg(jj \eb)$ \\ \cline{2-7}
& $P_e=0$ & $0.0766 $ & $4.746 $ & $4.373 $ & $0.806 $  & $2972$ \\ 
& $P_e=-80\%$ & $0.0906 $ & $6.826 $ & $5.940$ & $1.042 $ &  $3157 $ \\ \toprule
\end{tabular}
\caption{The SM cross sections in units of pb for the productions of $H$, $\wpm$, $Z$,
and two jets at the LHeC and FCC-he.
Here $\jf$ denotes the forward jet and $\eb$ the backward electron.
The values are based on the parton-level calculation in the four-flavor scheme for protons.
The kinematic phase space is constrained by
$p_T^{e,j}>10\gev$, $|\eta_{e,j}|<5$, and $\Dt R_{ej,jj} > 0.4$.
}
\label{table:Xsec:SM}
\end{center}
\end{table}

Before delving into the potential channels to discover the LFV signatures
at electron-proton colliders,
it is useful to summarize the total production cross sections of  the Higgs boson, $\wpm$, $Z$, 
and two QCD jets in the SM.
Table \ref{table:Xsec:SM} presents the cross sections at the LHeC and FCC-he, with the configurations of
\bea
\label{eq:configuration}
\hbox{LHeC:} && E_e=50\gev, \quad E_p=7\tev, %\; P_e = -80\%,
%\quad \int \mathcal{L} = 1\iab, 10\iab,
\\ \nn
\hbox{FCC-he:} && E_e=60\gev, \quad E_p=50\tev.%\; P_e = -80\%,
%\quad \int \mathcal{L} = 1\iab, 10\iab,
\eea
We used \textsc{MadGraph\_aMC@NLO}~\cite{Alwall:2011uj} with four-flavor scheme
and \texttt{NNPDF31\_lo} parton distribution
function  set for a proton~\cite{Ball:2017nwa}.
We classified the processes in two categories, the CC and NC processes.
In the CC process, a forward\footnote{Note that the direction of the proton beam is defined as forward.} jet $\jf$ accompanies a backward neutrino, 
while the NC process involves a backward electron $\eb$ and a forward jet $\jf$. 
Two choices for the electron beam polarization, $P_e=0$ and $P_e=-80\%$, are also considered.

In terms of production mechanisms, 
the Higgs boson, $\wpm$, and $Z$ undergo vector boson fusion, 
except for the NC production of the $Z$ boson.
As a result, the production cross sections are generically small,
around $\mco(1)\pb$ for the $\wpm/Z$ production and $\mco(0.1)\pb$ for the Higgs boson production at the LHeC.
Despite the higher proton beam energy at the FCC-he, 
there is no significant enhancement in the production cross sections 
due to the dependence on the square of the center-of-mass energy, $s=4E_eE_p$.
On the other hand, the electron beam polarization has a considerable impact on the production cross sections.
When $P_e=-80\%$, the CC production cross section experiences an enhancement by a factor of approximately $1.8$.
However, the increase in the NC production cross section is smaller.
Given the inherently small LFV signal rate, therefore, the CC production channel becomes 
 a more efficient choice to improve the sensitivity of the BSM search.
 
It is worth noting that the QCD multijet backgrounds of $jj\nu$ and $jj\eb$ do not give rise to a forward jet, 
because an additional jet radiates from the initial or final quark/gluon. 
Another unique feature of electron-proton colliders is the absence of prompt $W^+$ production in the CC channel.
Furthermore, in the case of single top quark production, only the anti-top quark can be produced through the CC process, and its decay products do not include a $W^+$ boson.
These unique features assist in suppressing the backgrounds for LFV signals involving the $Z$ and Higgs bosons.

\begin{figure}[!h]
    \centering
    \includegraphics[width=0.85\linewidth]{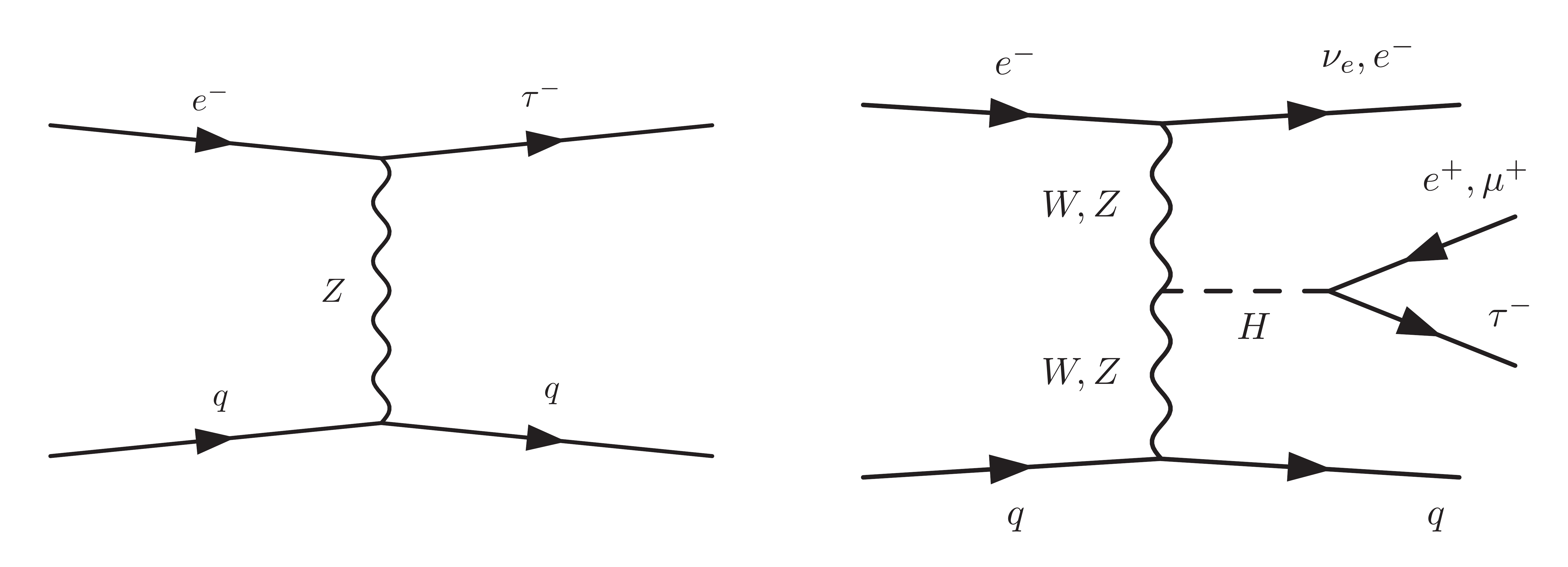}
    \caption{Feynman diagrams for the LFV signal of the  $Z$ boson (left panel)
    and the Higgs boson (right panel) at electron-proton colliders.
   }
    \label{fig-Feynman-Signal-Z}
    \end{figure}

We now  explore the potential channels for detecting the LFV phenomena in the $Z$ sector.
Unfortunately, conducting direct searches for the on-shell decay of $Z\to L_\al L_\bt$ 
is unfeasible at the LHeC and FCC-he.
At the LHeC,
the total production cross sections of a $Z$ boson even with $P_e = -80\%$ are
$824 \fb$ and $314\fb$  in the CC and NC channels, respectively. 
The extremely small branching ratios of $Z\to L_\al L_\bt$ listed in Table \ref{table:bounds} make 
it impossible to achieve observable event rates with the anticipated total luminosity of $1\iab$.

Fortunately, we can indirectly investigate the $Z$-$e$-$\tau$ vertex through the $t$-channel 
mediation of the $Z$ boson
by exploiting the initial electron beam.\footnote{The LFV of a $Z'$ boson through the $t$-channel at the LHeC
was studied in Ref.~\cite{Antusch:2020fyz}.
}
Our proposed process is as follows:
\begin{align}
p + e^- \to \jf + \tau^-.
\end{align}
The corresponding Feynman diagram is depicted in the left panel of \fig{fig-Feynman-Signal-Z}. 
The final state comprises one forward jet and one backward tau lepton with negative electric charge. 
The tau lepton can be identified if it decays hadronically
because the tau jet yields a fewer particle multiplicity and a localized energy deposit~\cite{CMS:2007sch,Bagliesi:2007qx,CMS:2018jrd}.
We will use the notation $\tauh$ for the hadronically decaying tau lepton in what follows.
The NC backgrounds are easily suppressed by vetoing the event with an electron.
The primary background originates from the QCD process $jj\nu_e$ in the CC mode,
where one jet is mistagged as $\tauh$.
The second dominant background arises from the CC process of $p e^- \to W^- +\jf \nu_e$, followed by $W^-\to \tau^-\nu_\tau$.
Additionally, we will consider the CC process of $Z(\to\ttau) +\jf \nu_e$,
where one tau lepton escapes detection.

In the study of LFV phenomena of the Higgs boson, we can consider two types of processes: 
direct decays $H \to e\tau/\mu\tau$
and an indirect process $e^- p \to \tau^- j$
mediated by the Higgs boson in the $t$-channel.
However, the indirect process is subject to significant suppression 
due to the extremely small Yukawa couplings between the Higgs boson and leptons, 
as well as the loop-induced couplings of the Higgs boson to gluons. 
As a result, our primary focus is directed towards the on-shell decays of the Higgs boson.

For the decay of $H\to e\tau$,
let us first consider the CC production of the Higgs boson.
As shown in Table \ref{table:Xsec:SM},
the CC production of $W^+$,
which could yield the significant background to the positron final state, vanishes at the leading order.
Therefore, we propose the signal process of
\bea
\label{eq:signal:H:etau}
p + e^- \to H(\to e^+ \tau^-) + \jf\nu_e ,
\eea
of which the final state consists of a positron,
a negatively charged $\tauh$, a forward jet, and a backward neutrino.
The Feynman diagram is in the right panel of \fig{fig-Feynman-Signal-Z}.
Two backgrounds contribute.
The primary one arises from $p e^- \to Z +\jf\nu_e $ followed by $Z\to \tau_{e^+} \tauh^-$
or $Z \to j_{e^+} j_{ \tauh}$.
Here $\tau_{e^+}$ represents the tau lepton decaying into $\tau^+ \to e^+ \nu_e \bar{\nu}_\tau$,
and $j_X$ is a QCD jet mistagged as the particle $X$.
The second dominant background is associated with the production of multiple jets and a neutrino,
$p e^- \to j_{e^+} j_{ \tauh} +\jf \nu_e$,
where one jet is mistagged as a positron and the other jet is mistagged as $\tauh$.

To explore the NC process of $H\to e\tau$, we propose incorporating both $H\to e^+ \tau^-$ and $H\to e^-\tau^+$  due to
the small NC production cross section. The suggested process is as follows:
\bea
p + e^- \to H(\to e^\pm \tau^\mp) + \jf\eb .
\eea
In this case, the final state consists of an electron/positron, $\tauh$,  a forward jet, and the backward electron. 
The primary background stems from $pe^- \to Z +\jf \eb $, followed by $Z\to \tau_e\tauh$
or $Z\to j_e j_\tauh$.
Another significant backgrounds arise from the NC processes of $p e^- \to \wpm (\to e^\pm\nu) j_\tau +\jf\eb$
and  $p e^- \to \wpm (\to j_e j_\tau) +\jf\eb$.  
Lastly, there is a background from $pe^- \to H + \jf\eb$, followed by $H \to \tau_e\tauh$.
The backgrounds originating from the CC processes can be efficiently suppressed 
by vetoing events with missing transverse energy ($\met$) and requiring a backward electron.

For the decay of $H\to \mu\tau$ in the CC mode,
we consider the final state consisting of a muon (regardless of its electric charge), a tau lepton, the missing transverse energy,
and a forward jet.
As there is no electron  in the signal event,
the NC background processes are easily controlled.
Among the CC processes,
the dominant background originates from $p e^- \to Z +\jf\nu_e $ followed by $Z\to \tau_{\mu} \tauh$.
The second background arises from $pe^- \to H +\jf \nu $, followed by $H \to \tau_\mu\tauh$.
In the NC signal process of $H\to \mu\tau$, the final state comprises a muon, a tau lepton,
a backward electron, and a forward jet.
The main background is from $pe^- \to Z +\jf\eb $, followed by $Z\to \tau_\mu\tauh$.
The subleading background comes from $pe^- \to H +\jf\eb $, followed by $H \to \tau_\mu\tauh$.

\begin{table}
%\begin{ruledtabular}
  {\renewcommand{\arraystretch}{1.2} 
  \begin{tabular}{|c|c|c|c|c|}
\toprule
\multicolumn{3}{|c|}{} & Signal & Backgrounds\\ \hline 
LFV $Z$ & $e\tau$ &NC & $\jf \tau^-$ & $ W^-(\to \tau^- \nu)\big/Z(\to \tau\tau_{\rm un})+\jf\nu$, ~$jj_\tauh\nu$
\\ \hline
\multirow{5}{*}{~LFV $H$~} & \multirow{3}{*}{~$e\tau$~} & ~CC~ & ~$H(\to e^+\tau^-)+\jf\nu$~ & $Z(\to \tau_{e^+}\tauh )+\jf\nu$, $ j_{e} j_\tauh +\jf \nu$
\\ \cline{3-5}
& & \multirow{2}{*}{~NC~} & \multirow{2}{*}{~$H(\to e^\pm\tau^\mp)+\jf\eb$} & 
~~~$Z(\to \tau_e\tauh)\big/\wpm(\to j_e j_\tauh)\big/H(\to \tau_e\tauh)+\jf\eb$~~~ \\ \cline{5-5}
& & & & $\wpm (\to e^\pm\nu) j_\tau +\jf\eb$ \\ \cline{2-5}
& \multirow{2}{*}{~$\mu\tau$~} & ~CC~ & ~$H(\to \mu^\pm\tau^\mp)+\jf\nu$~ & $Z(\to \tau_{\mu}\tauh )\big/H(\to \tau_\mu\tauh)+\jf\nu$
\\ \cline{3-5}
& & ~NC~ & ~$H(\to \mu^\pm\tau^\mp)+\jf\eb$ & $Z(\to \tau_{\mu}\tauh )\big/H(\to \tau_\mu\tauh)+\jf\eb$
\\ \bottomrule
\end{tabular}
}
\caption{\label{table:signal:BG}
The signal and background processes of the LFV phenomena of the $Z$ and Higgs bosons
at electron-proton colliders.
Here $\tau_\ell$ denotes the tau lepton decaying into $\ell\nu_\ell\nu_\tau$ ($\ell=e,\mu$),
$\tauh$ is the hadronically decaying $\tau$,
$j_X$ is the QCD jet mistagged as the particle $X$,
and $\tau_{\rm un}$ is the tau lepton escaping the detection.
}
\end{table}

In Table \ref{table:signal:BG}, we summarize the potential discovery channels to probe the LFV phenomena
of the $Z$ and Higgs bosons
along with the corresponding backgrounds.
Regarding the LFV of the $Z$ boson, the only feasible option at electron-proton colliders is the indirect probe 
through the process $p e^- \to \jf\tau^-$.
For the LFV in the $H$ sector,
we suggest two decay modes of $H\to e^+\tau^-$ and $H\to \mu^\pm\tau^\mp$
in the CC production of the Higgs boson because
the NC mode has a few disadvantages.
First,  the NC production cross section of the Higgs boson is merely about 10\% of the CC production cross section, 
as shown in Table \ref{table:Xsec:SM}.
Second,
the electron beam polarization of $P_e=-80\%$ diminishes the discovery potential
since the background
cross sections of the NC production  of $Z$ and $\wpm$ increase more than that of $H$.

Now we discuss the parton-level cross sections of the signals and backgrounds.
For the signal, we first obtained the Universal FeynRules Output (UFO)~\cite{Degrande:2011ua} file 
for the BSM with the Lagrangian in \eq{eq:lg:Z} and \eq{eq:lg:H}.
The event generation at leading order is performed 
using \textsc{MadGraph\_aMC@NLO}~\cite{Alwall:2011uj}  version 3.4.2 with \texttt{NNPDF31\_lo} parton distribution
function  set~\cite{Ball:2017nwa}.
The generator-level cuts were imposed on the parton-level 
objects like $p_T^j > 10~$GeV, $\Delta R_{ij} > 0.4$, and $|\eta_j| < 5$.
Here $\Dt R = \sqrt{ \lf \Dt \eta\ri^2 +  \lf \Dt\phi\ri^2 }$
with $\eta$ and $\phi$ being the rapidity and azimuthal angle, respectively.
For the renormalization and factorization scales,
we take 
\bea
\label{eq:scales}
\mu_{R,0} = \mu_{F,0} \equiv \frac{1}{2} \sum_i \sqrt{p_{T,i}^2 + m_i^2},
\eea
where $i$ covers all the particles in the final state.

By setting $C_{\tau e}^L = C_{\tau e}^R =10^{-3}$,
we calculated the parton-level cross sections of $p e^- \to \jf \tau^-$ 
as
\bea
\hbox{LHeC: } & \sg(p e^- \to \jf \tau^-)\Big|_{C_{\tau e}^{L,R}=10^{-3}} = \left\{
\begin{array}{ll}
1.29\fb & \hbox{ for } P_e=0; \\
1.31\fb & \hbox{ for } P_e=- 80\%; \\
\end{array}
\right.
\\[7pt] \nn
\hbox{FCC-he: } & \sg(p e^- \to \jf \tau^-)\Big|_{C_{\tau e}^{L,R}=10^{-3}}= \left\{
\begin{array}{ll}
3.38\fb & \hbox{ for } P_e=0; \\
3.41\fb & \hbox{ for } P_e=- 80\%. \\
\end{array}
\right.
\eea
It is evident that the signal cross section is not significantly enhanced by $P_e = -80\%$.
Since the cross sections for the main CC backgrounds  increase by a factor of about 1.8,
 we set $P_e=0$ for the LFV studies of $Z$ . 
For the LFV phenomena of the Higgs boson, 
the signal cross sections of $p e^- \to H (\to e\tau/\mu\tau)+\jf \nu$
can be obtained by multiplying the branching ratio and the Higgs production cross sections presented in Table \ref{table:Xsec:SM}. 
If $\br(H  \to e\tau/\mu\tau) =10^{-3}$,
the total cross section of the signal with $P_e=-80\%$ is about $145\ab$ at the LHeC and about $604\ab$ at the FCC-he.

\section{Signal-to-background analysis}
\label{sec:analysis}
%%%%%%%%%%%%%%%%%%%%%%%%%%%%%%%%%%%%%%%%%%%%%%%%%%%%%
In this section,
we conduct a comprehensive analysis of the signal and backgrounds through the full simulation at the detector level.
The Monte Carlo event generation procedure is as follows.
First, we employ \textsc{MadGraph\_aMC@NLO}  
to generate events for both the signal and backgrounds at leading order
with parton distribution functions using the \texttt{NNPDF31\_lo} PDF set and $\alpha_s(m_Z^2) = 0.118$.
%Subsequently, we convolve the parton-level cross-sections with the PDFs in the \texttt{LHAPDF6} library,
%assuming that the direction of the proton beam is forward.
To accurately simulate the decays of  $\wpm$ and $Z$, as well as to account for parton showering and hadronization effects, 
we use \textsc{Pythia} version \texttt{8.309}~\cite{Bierlich:2022pfr}.
In order to incorporate more stable performance of the \textsc{Pythia} for electron-proton colliders,
we made some modifications to the default setup, setting
\texttt{partonlevel:mpi=off},
\texttt{SpaceShower:dipoleRecoil=on},
\texttt{PDF:lepton=off}, and 
\texttt{TimeShower:QEDshowerByL=off}.

%We make specific modifications to the default setup of \textsc{Pythia8} to ensure proper modeling of hadronization optimized for electron-proton colliders.  
%These modifications include deactivating the lepton PDF, disabling QED initial state radiation for the electron beam, and turning off multiple-parton interactions.

%LHeC with $P_e=-80\%$

To simulate the detector effects, we utilized version 3.5.0 of \textsc{Delphes}~\cite{deFavereau:2013fsa}. 
We adjust the \textsc{Delphes} card\footnote{The \textsc{Delphes} cards specialized for the LHeC
and FCC-he can be found in the GitHub repository \url{https://github.com/delphes/delphes/tree/master/cards}}
to align the particle efficiencies, momentum smearing, and isolation parameters with the default values specified in the LHeC Concept Design Report~\cite{Agostini:2020fmq}.
Jet clustering was performed using the anti-$k_T$ algorithm~\cite{Cacciari:2008gp} 
with a jet radius $R=0.4$ implemented in \texttt{FastJet} version 3.3.4~\cite{Cacciari:2011ma}.

The $\tauh$-tagging plays a crucial role in detecting the LFV signals of the $Z$ and Higgs bosons.
Let us elaborate on our strategy for choosing the $\tauh$-tagging efficiency in more detail.
The $\tauh$-tagging efficiency is not fixed but rather a parameter that we can choose depending on our target process.
When selecting the $\tauh$-tagging efficiency, it is essential to strike a balance between retaining the $\tau$-related signal events and suppressing the background events.
To find this balance, we take into account the feature that as the tagging efficiency increases, the mistagging rate also increases.

For the LFV $Z$ channel, the primary background is $jj\nu$.
Given that this background overwhelmingly dominates the signal, our primary objective is to effectively suppress the QCD background. 
To achieve this, we chose a relatively small $\tauh$-tagging efficiency. 
By doing so, we reduce the mistagging probability, thereby minimizing contamination from QCD jets and enhancing the sensitivity to the LFV $Z$ signal.
 In our analysis, 
 we have specifically chosen $P_{\tauh\to\tauh}=0.4$ for the $\tauh$-tagging efficiency and $P_{j\to \tauh}=0.001$ for the mistagging probability. 
 These values are based on the default settings in the \textsc{Delphes} cards for the LHeC and FCC-he.

On the other hand, in the LFV $H$ channel, the dominant background arises from 
$p e^- \to Z (\to \tau_{e,\mu}\tau) + \jf \nu$ processes.
Unlike the LFV $Z$ channel, this background does not overwhelm the signal. 
Consequently, choosing a small $\tauh$-tagging efficiency would not be advantageous since it would decrease both the signal and background event yields. 
To maximize the signal significance in the presence of a manageable background,  
we opt for a larger $\tauh$-tagging efficiency. This choice allows us to retain more $\tau$-related signal events while still keeping the background under control.
For the LFV $H$ analysis, we have adopted $P_{\tauh\to\tauh}=0.85$ for the $\tauh$-tagging efficiency and $P_{j\to \tauh}=0.05$~\cite{ATLAS:2022aip}.

We compute the signal significance $\mcs$  taking into account the uncertainty in the background,
defined by~\cite{Cowan:2010js}
\bea
\label{eq:significance:nbg}
\mathcal{S} = 
\left[2(\nsg + \nbg) \log\left(\frac{(\nsg + \nbg)(\nbg + \dbg^2)}{\nbg^2 + (\nsg + \nbg)\dbg^2} \right)  
- 
\frac{2 \nbg^2}{\delta_b^2} \log\left(1 + \frac{\dbg^2 \nsg}{\nbg (\nbg + \dbg^2)}\right)\right]^{1/2},
\eea
where $\nsg$ is the number of signal events, $\nbg$ is the number of background events, 
and $\dbg = \Delta_{\rm bg} \nbg$ is the uncertainty in the background yields.

\subsection{Results of the LFV $Z$}
\label{sec:analysis:Z}
%%%%%%%%%%%%%%%%%%%%%%%%%%%%%%%%%%%%%%%%%%%%%%%%%%%%%

For a benchmark point, we consider the following values:
\begin{align}
C_{\tau e}^L = C_{\tau e}^R = 10^{-3}, \quad P_e=0.
\end{align}
Focusing on the LHeC first,
we generated $5\times 10^5$ events for the signal, $3.25\times 10^7$ for $jj\nu$,
 $1.25 \times 10^7$ for $W^- + \jf \nu$, and  $1.25 \times 10^7$ for $Z + \jf \nu$
 at the \textsc{Madgraph} level.
To account for the detector effects, we simulated the events using \textsc{Pythia} and \textsc{Delphes}.
Afterwards, we applied the basic selection criteria, which require at least one $\tauh$ jet and one QCD jet (i.e., non-$\tauh$ jet) 
with transverse momentum $p_T^{\tauh,j} > 20\gev$ and pseudorapidity $|\eta_{\tauh,j} | < 5$.
We repeated the same procedure
 for the FCC-he.

\begin{figure}[!t]
    \centering
    \includegraphics[width=0.95\linewidth]{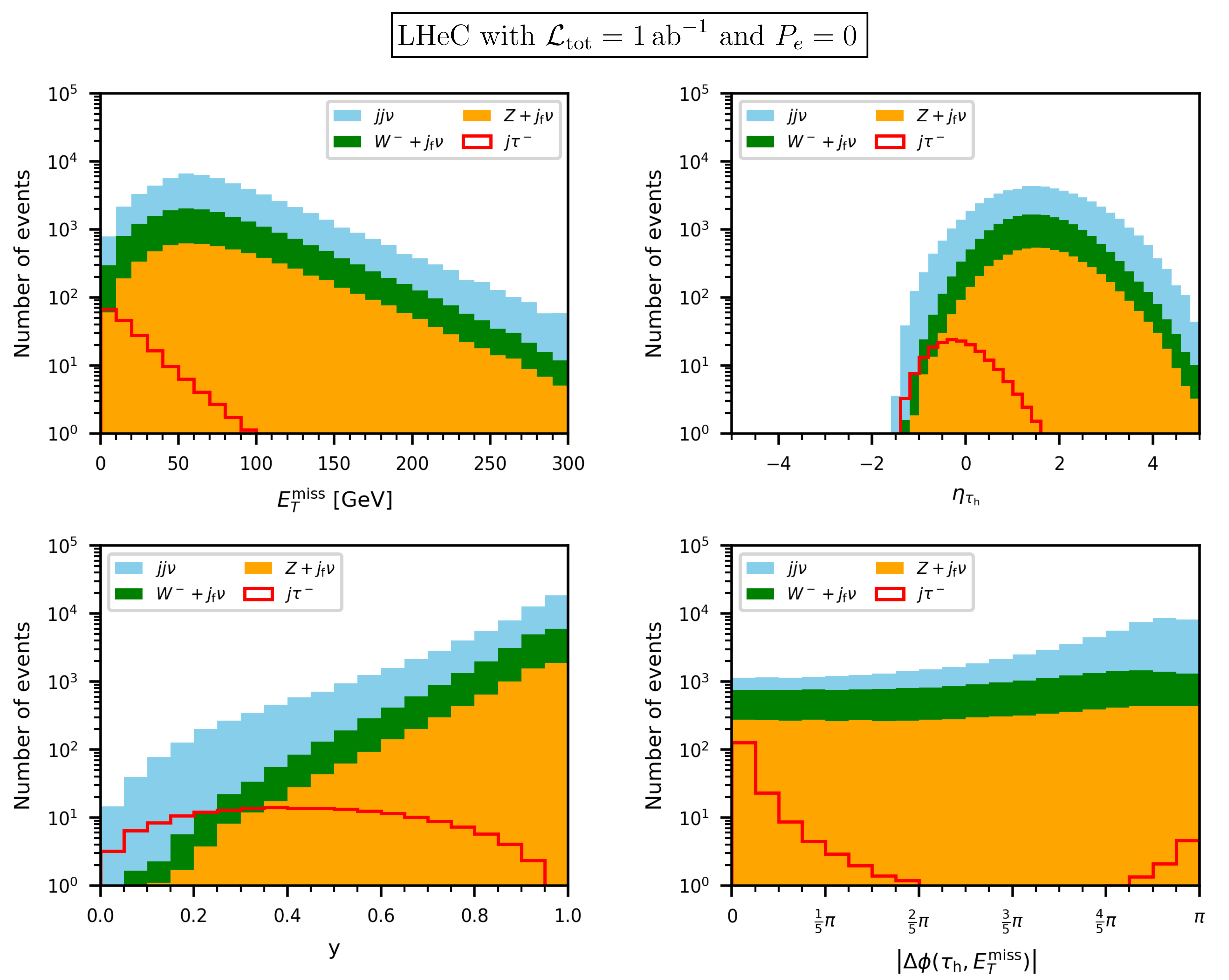}
    \caption{Kinematic distributions for the signal of $pe^- \to  \jf \tau^-$ and the corresponding backgrounds
     about missing transverse energy 
    $\met$ (top-left), $\eta_\tauh$ (top-right), $y$ (bottom-left), 
    $|\Dt \phi(\tauh,\met)|$ (bottom-right) at the LHeC with the total integrated luminosity of $\mathcal{L}_\tot = 1\iab$ and $P_e=0$.
    The signal is for $C_{\tau e}^L = C_{\tau e}^R = 10^{-3}$.
    The results are based on the dataset after the basic selection described in the text. 
The various background contributions are overlaid, creating a stacked representation, 
while the anticipated signal is depicted by the red solid line.    }
    \label{fig-kin-distribution-Z}
    \end{figure}

In order to devise an optimal strategy for the cut-based analysis, it is crucial to examine various kinematic distributions. 
In \fig{fig-kin-distribution-Z},
we present selected distributions after applying the basic selection criteria at the LHeC\footnote{The kinematic distributions observed at the FCC-he closely resemble those obtained at the LHeC.
}
with $\lumtot = 1\iab$ and $P_e=0$. 
The contributions from different backgrounds are stacked on top of each other, 
while the expected signal for $C^{L/R}_{\tau e} = 10^{-3}$ is depicted by the red solid line.
The kinematic variables for the distributions include 
    $\met$ (top-left), $\eta_\tauh$ (top-right), $y$ (bottom-left), and
    $|\Dt \phi(\tauh, \met) |$ (bottom-right).
    Here $y$ is one of the variables used in Deep Inelastic Scattering,
    defined as
    \bea
    \label{eq:y}
    y = \frac{p_p \cdot (p_e - p_\tauh)}{p_p \cdot p_e},
    \eea
    where $p_p$ represents the four-momentum of the proton beam.
For  the azimuthal angle difference, $\left| \Dt \phi(\tauh, \met) \right|$,
we take the absolute value to cover the physical range $[0,\pi]$ from collinear to back-to-back configurations.

The distribution of $\met$ exhibits a distinct difference between the signal and backgrounds.
In the signal, it shows a monotonic decrease since $\met$ is solely attributed to the neutrino from $\tau\to\tauh\nu_\tau$ decays.
In contrast, the CC backgrounds, which involve the production of a prompt neutrino, display a bump structure in the $\met$ distribution,
peaked around $\met \sim 65\gev$.
Therefore, applying an appropriate upper cut on $\met$ would be an effective strategy.
Furthermore, the distribution of $\eta_\tauh$ reveals a noticeable discrepancy between the signal and backgrounds.
In the signal,
the $\tauh$-jet originating directly from the electron beam is predominantly observed in the backward region.
On the other hand, the backgrounds tend to produce more forward $\tauh$ due to the higher energy of the proton beam 
compared to the electron beam.

The distributions of $y$ also yield a distinct pattern between the signal and backgrounds.
The signal distribution shows a relatively uniform spread across the range $y\in[0.05, 0.9]$,
while the background distributions exhibit a peak around $y\simeq 1$.
This difference can be understood by considering the definition of $y$ in the proton rest frame,
$y = (E_e^{\rm rest} - E_\tauh^{\rm rest})/E_e^{\rm rest} $.
In the signal events, which predominantly produce backward $\tauh$ in the laboratory frame,
the Lorentz boost into the proton rest system can result in various values of $E_\tauh^{\rm rest}$ and thus the flat distribution in $y$.
On the other hand, the backgrounds mostly feature forward $\tauh$ in the lab frame. 
In the proton rest frame, the $\tauh$ almost comes to a stop, resulting in $E_\tauh^{\rm rest} \simeq 0$ and consequently $y\simeq 1$. 

Let us delve into the distribution of the azimuthal angle difference between $\tauh$ and the missing transverse momentum $\met$.
In the signal, $\left| \Dt \phi(\tauh,\met)\right|$ is peaked around zero.\footnote{We also observe a smaller peak around  $|\Dt \phi(\tauh, \met)| \simeq \pi$ in the signal.
This feature emerges due to the adoption of a relatively lower tagging efficiency $P_{\tauh\to\tauh} =0.4$
and the requirement of having one $\tauh$ and one QCD jet in the basic selection.
In the signal events with $|\Dt \phi(\tauh, \met)| \simeq \pi$,
the object initially tagged as $\tauh$  is a QCD jet,
and the object initially tagged as $j$ is $\tauh$ in fact,
as confirmed by $|\Delta\phi(j, \met)| \simeq 0$.
}
This behavior arises because $\met$ solely originates from $\tau\to \tauh\nu_\tau$,
which naturally leads to the collinear motion of $\tauh$ and $\met$.
For the backgrounds, on the contrary,
$\met$ primarily originates from prompt neutrinos.
 As a result, the backgrounds exhibit a larger angular separation between $\tauh$ and $\vmet$ compared to the signal events.
 
\begin{table*}[!t]
\setlength\tabcolsep{10pt}
\centering
{\renewcommand{\arraystretch}{1.1} 
\begin{tabular}{|c||c|c|c|c||c|c|}
\toprule
\multicolumn{7}{|c|}{Cross sections in units of fb at the LHeC with  $P_e=0$ }\\ \toprule
Cut    &  Signal & $jj\nu$ & $W^- + \jf\nu$  & $Z+\jf\nu$ & $\mcs_{1\iab}^{10\%}$ &  $\mcs_{3\iab}^{10\%}$  \\[3pt] \hline
Basic                                                & 0.1854 & 40.3044 & 13.5870 & 6.4368 & 0.03 & 0.03\\ \hline
$Q(\tauh)<0$                                         & 0.1852 & 20.7435 & 12.9228 & 3.2324 & 0.05 & 0.05\\ \hline
$\met<10\gev$                                        & 0.0668 & 0.2654 & 0.2207 & 0.0295 & 1.14 & 1.21\\ \hline
$\eta_\tauh<0$                                       & 0.0512 & 0.1293 & 0.0164 & 0.0031 & 2.45 & 2.84\\ \hline
$ y < 0.5$                                           & 0.0465 & 0.0761 & 0.0048 & 0.0012 & 3.39 & 4.18\\ \hline
$ \left| \Dt \phi(\tauh, \met) \right| < \pi/10$      & 0.0261 & 0.0051 & 0.0007 & 0.0001 & 7.08 & 11.17\\ \hline
\toprule
\multicolumn{7}{|c|}{Cross sections in units of fb at the FCC-he with $P_e=0$}\\ \toprule
Cut    &  Signal & $jj\nu$ & $W^- + \jf\nu$  & $Z+\jf\nu$ & $\mcs_{1\iab}^{10\%}$ &  $\mcs_{10\iab}^{10\%}$  \\[3pt] \hline
Basic                                                 & 0.4904 & 165.6040 & 75.1131 & 37.0154 & 0.02 & 0.02 \\ \hline
$Q(\tauh)<0$                                          & 0.4895 & 86.8901 & 71.2757 & 18.4660 & 0.03 & 0.03 \\ \hline
$\met<10\gev$                                         & 0.1454 & 0.8508 & 0.7763 & 0.1089 & 0.79 & 0.81 \\ \hline
$\eta_\tauh<0$                                        & 0.1165 & 0.4066 & 0.0770 & 0.0127 & 2.00 & 2.17 \\ \hline
$y<0.3$                                               & 0.0856 & 0.1606 & 0.0068 & 0.0024 & 3.56 & 4.27 \\ \hline
$\left| \Dt \phi(\tauh, \met) \right| < \pi/10$       & 0.0470 & 0.0137 & 0.0010 & 0.0004 & 8.04 & 15.32 \\ 
\bottomrule
\end{tabular}
}
\caption{Cut-flow chart of the cross sections of the signal with $C_{\tau e}^L =C_{\tau e}^R=10^{-3}$ 
and the corresponding backgrounds at the LHeC and 
the FCC-he with the unpolarized electron beam. 
For the significance $\mcs$,
two cases of the total luminosity are considered,
$1\iab$ and $3\iab$ for the LHeC
while $1\iab$ and $10\iab$ for the FCC-he.
We include a 10\% background uncertainty.
}\label{tab:Z:cutflow}
\end{table*}

Table \ref{tab:Z:cutflow} illustrates the cut-flow chart, presenting the cross sections of the signal and backgrounds 
in units of fb at the LHeC and FCC-he, assuming an unpolarized electron beam.  
The signal significances are provided, taking into account a 10\% background uncertainty ($\Delta_{\text{bg}} = 10\%$), 
for two reference values of the total integrated luminosity:
$1\iab$ and $3\iab$ at the LHeC,\footnote{While the CDR of the LHeC specifies a total luminosity of $1\iab$, 
we additionally calculated the significance for $\lumtot = 3\iab$ to demonstrate the potential gains achievable by tripling the luminosity.} and $1\iab$ and $10\iab$ at the FCC-he.

Each kinematic cut plays a crucial role in distinguishing the signal from the backgrounds. 
The most influential discriminator is $\met<10\gev$,
which increases the significance by approximately a factor of 20.
By applying a series of kinematic cuts on $\eta_\tauh$, $y$, and $|\Delta \phi(\tauh, \met)|$, the signal significance is further enhanced.
Even with a total luminosity of $\lumtot = 1\iab$, 
the LFV $Z$ couplings of $C_{\tau e}^L = C_{\tau e}^R = 10^{-3}$ 
can be measured with a signal significance of about 7.08 (8.04) at the LHeC (FCC-he).
Increasing the LHeC (FCC-he) luminosity into $3\iab$ ($10\iab$)
will improve the significance into $\mcs^{10\%} \simeq 11.2~(15.3)$.

To explore the maximum sensitivity of the LHeC and FCC-he in probing the LFV $Z$, 
we extend our analysis by employing a multivariate analysis approach.
We utilized the boosted decision tree (BDT) algorithm~\cite{Roe:2004na}, 
harnessed through the highly effective Extreme Gradient Boosting (\textsc{XGBoost}) package~\cite{Chen:2016btl}.
\textsc{XGBoost} is one of the most powerful machine learning algorithms,
showing its superiority over other methods, 
such as Deep Neural Networks and Support Vector Machines. 
Its exceptional performance can be primarily attributed 
to its unique ensemble learning techniques, 
most notably gradient boosting. 
The remarkable success of \textsc{XGBoost} is exemplified 
by its triumphs in esteemed Kaggle competitions~\cite{kaggle1, kaggle2}.
Within the particle physics community, \textsc{XGBoost} has recently gained popularity and is actively being applied in various contexts, including the analysis of the SM Higgs boson~\cite{ATLAS:2017ztq, CMS:2020tkr, ATLAS:2021ifb, CMS:2020cga, CMS:2021sdq}, dark matter~\cite{ATLAS:2021jbf}, vector quarks~\cite{Dasgupta:2021fzw}, a composite pseudoscalar~\cite{Cornell:2020usb}, and the new strategies for faster event generation~\cite{Bishara:2019iwh}.
In our study, we utilized \textsc{XGBoost} as a binary classifier. 

For training the model,
we generated $5 \times 10^5$ signal events, $3.25\times 10^7$ for $jj\nu$,
 $1.25 \times 10^7$ for $W^- + \jf \nu$, and  $1.25 \times 10^7$ for $Z + \jf \nu$
 at the \textsc{Madgraph} level.
The dataset was then processed through \textsc{Pythia} and \textsc{Delphes}.
After applying the basic selection, we divided the dataset into three parts, 52.5\% for training, 30\% for testing, and 17.5\% for validating the algorithm.
We use the following 11 observables as inputs for the BDT analysis:
\bea
\label{eq:BDT:Z:variables}
&& \met, \quad \sqrt{-q^2}, \quad \eta_\tauh, \quad \eta_j, \quad
p_T^\tauh,  \quad  p_T^j,  \\ \nn &&
|\Dt \phi(\tauh, j)|,  \quad |\Dt \phi(\tauh, j)|, \quad
\Dt R(\tauh, j), \quad x,\quad y,
\eea
where $q^2$ and $x$ are defined by
\bea
\label{eq:qsq:x}
q^2 = (p_e - p_\tauh)^2, \quad x  = -\frac{q^2}{2 p_p \cdot  (p_e - p_\tauh)} \;.
\eea

We implemented several strategies to enhance the performance of the BDT.
Firstly,  we used the area under the receiver operating characteristic curve to measure the BDT performance
because there exists a significant imbalance between the signal and background samples.
Secondly, we addressed the issue of determining the optimal stopping point for the boosting rounds 
to prevent overfitting and ensure the generalization of the model to unseen test samples.
To accomplish this, we incorporated a validation set and continuously monitored its performance during the training process. 
 If a noticeable decline in the validation performance was observed, it served as an indicator to stop the training.
In our implementation, we set \texttt{early\_stopping\_rounds=3}.
For the BDT output, we used the default settings in \textsc{XGBoost} for binary classification tasks, 
employing the \texttt{binary:logistic} objective function, which represents the predicted probabilities of the positive class.

\begin{figure}[!t]
    \centering
    \includegraphics[width=\linewidth]{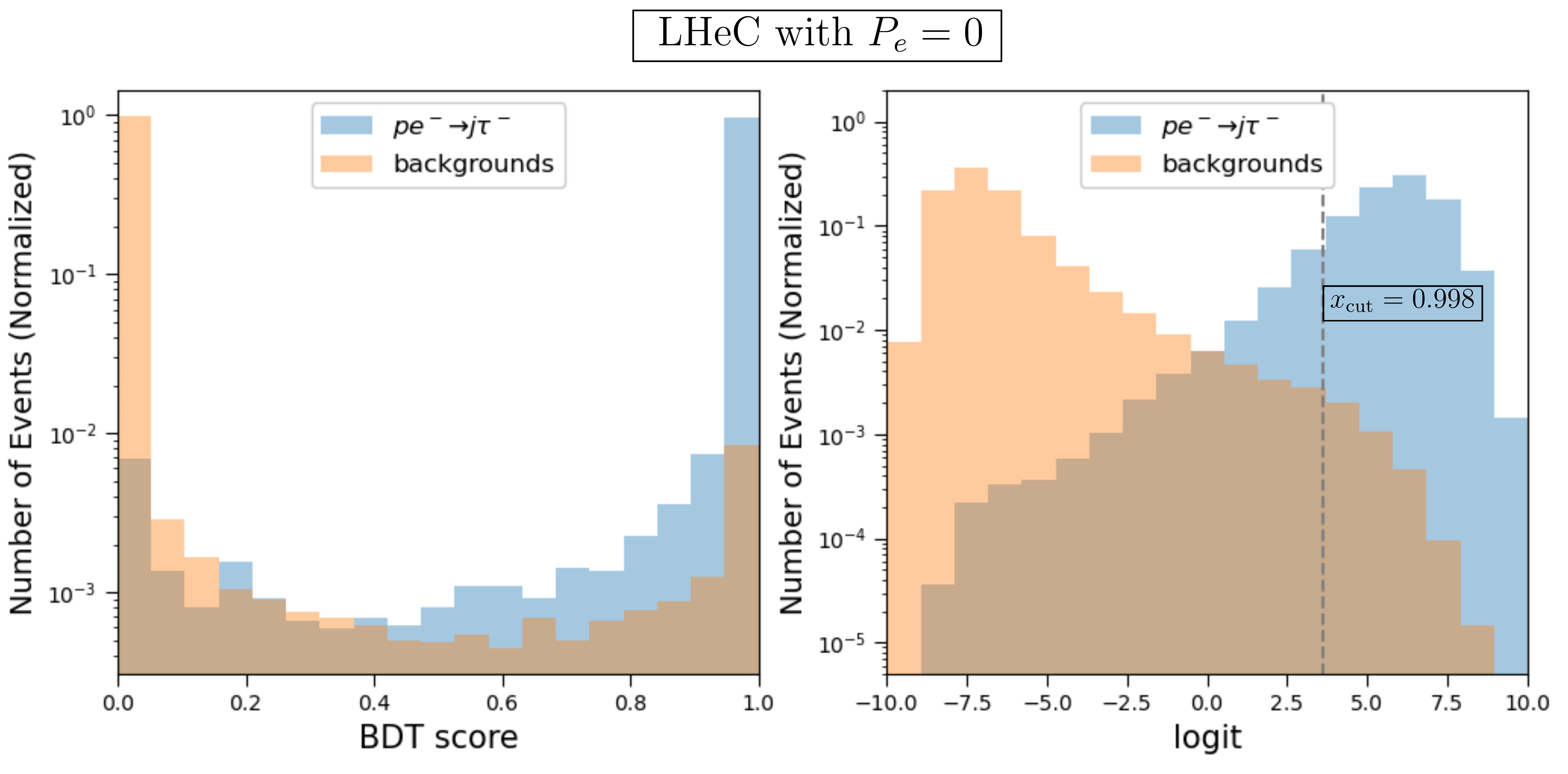}\\
    \caption{Normalized signal and background distributions
    against the BDT score (left panel) 
 and the raw untransformed \texttt{logit} values in \textsc{XGBoost}  for the signal process of $p e^- \to \jf \tau^-$.
    We consider the LHeC with  $P_e =0$.
The results are based solely on the testing dataset after the basic selection described in the text. 
}
    \label{fig-Z-BDT}
    \end{figure}

In \fig{fig-Z-BDT},
we  present
 the normalized distributions of the signal and backgrounds against the BDT score (left panel) 
 and the \emph{raw} untransformed \texttt{logit} values in \textsc{XGBoost} 
  for the signal process of $p e^- \to \jf \tau^-$
 at the LHeC with $P_e =0$.
We show the results exclusively from the testing dataset, which remains unseen by the model during the training and validation stages.
The distribution against the BDT scores clearly shows a distinct separation between the signal and backgrounds.
Furthermore, we found that the results at the FCC-he are similar to those at the LHeC.

\begin{table}
%\begin{ruledtabular}
  {\renewcommand{\arraystretch}{1.2}
  \begin{tabular}{|c|c|c|c|c||c|c|c|c|c|}
\toprule
\multicolumn{10}{|c|}{$p e^- \to \tau^- \jf$ via the LFV $Z$ with $C^L_{\tau e} = C^R_{\tau e} = 10^{-3}$}\\ \toprule
 \multicolumn{5}{|c||}{LHeC with $P_e=0$} & \multicolumn{5}{c|}{FCC-he with $P_e=0$} \\ \hline
$\lumtot$ & $\xc$ & $\nsg$ & $\nbg$  & $\mcs^{10\%}$ &  $\lumtot$ & $\xc$ & $\nsg$ & $\nbg$  & $\mcs^{10\%} $ \\ \hline
 \multirow{2}{*}{~~$1\iab$~~}  & ~~0.997~~    & ~~104.4~~   & ~~24.3~~  &  ~~12.3~~
 	&   \multirow{2}{*}{~~$1\iab$~~}  & ~~0.998~~   & ~~202.5~~    & ~~71.0~~  & ~~12.4~~ \\ \cline{2-5}\cline{7-10}
   & 0.900   & 192.2    & 405.0     & 3.76     &   & 0.900   & 494.8   & 1822.5  & 2.44 \\ \hline
 \multirow{2}{*}{$3\iab$}  & 0.997   & 313.2    & 73.0     & 16.8   &   \multirow{2}{*}{$10\iab$} & 0.999   & 1001.8   & 194.8  & 22.4 \\ 
 \cline{2-5}\cline{7-10}
  & 0.900   & 576.5 & 1214.9 & 4.01   &  & 0.900   & 4948.5   & 18224.6 & 2.50 \\ 
\bottomrule
\end{tabular}
}
\caption{\label{table:Z:significance}
The results of the BDT analysis
for the signals of $p e^- \to \tau^- \jf$ via the LFV $Z$ with $C^L_{\tau e} = C^R_{\tau e} = 10^{-3}$
at the LHeC and FCC-he with $P_e=0$, including 10\% background uncertainty.
Here $\nsg$ ($\nbg$) is the number of signal (background) events and
$\xc$ is the cut on the BDT output of the testing dataset.
}
\end{table}

In Table \ref{table:Z:significance}, 
we present
the results of the BDT analysis
for the signals of the LFV $Z$ at the LHeC and FCC-he with $P_e=0$ and a 10\% background uncertainty.
Two cases of the total integrated luminosity are considered,
$1\iab$ and $3\iab$ for the LHeC while $1\iab$ and $10\iab$ for the FCC-he.
The table includes the BDT score cut $\xc$,
the number of the signal events $\nsg$,
and the number of the total backgrounds $\nbg$.
The BDT score cut $\xc$ is selected to maximize the signal significance~\cite{Kang:2017rfw,Antusch:2020fyz,Wang:2020ips,Adhikary:2020cli} while ensuring that at least 10 background events survive.
The results clearly demonstrate the improvement achieved through the multivariate approach.
The significance increases by about 74\% (54\%) at the LHeC (FCC-he) with $\lumtot =1\iab$.
To showcase the performance of our chosen $\xc$, we also provide the results for a reference value of $\xc=0.9$ in the table. This additional information allows readers to better understand the effect of different BDT score cuts and provides a basis for comparison to assess the robustness of our analysis.

Before proceeding to the next topic, it is essential to address potential concerns about extreme values of $\xc$.
The BDT scores are obtained by applying the Softmax function to the raw \texttt{logit} scores. 
Consequently, judging the validity based solely on the face-value of the BDT score might not be reasonable.
Moreover, it is important to note that the sharp distribution of the BDT output is a characteristic feature of the Softmax function.
For example, setting a BDT cut value of $\xc = 0.999$ corresponds to a signal efficiency of approximately 20\%.

To better understand the underlying distributions, 
we have also included the distributions of the raw untransformed \texttt{logit} values in the right panel of \fig{fig-Z-BDT}. 
These values can be obtained by setting the objective to \texttt{binary:logitraw}.
By doing so, we can assess the true characteristics of the \texttt{logit} scores and draw more meaningful conclusions.
In the plot, the vertical dashed line corresponds to the BDT score cut $\xc = 0.998$, which not only demonstrates a satisfactory signal efficiency but also exhibits a moderate \texttt{logit} value.

Let us derive upper bounds on $\br(Z \to e^\pm \tau^\mp) $ based on our results.
Although $pe^- \to \jf \tau^-$ is sensitive only to $C^L_{\tau e}$ and $C^R_{\tau e}$, 
we can convert the results of $C^{L,R}_{\tau e}$ into the limits on $\br(Z \to e^\pm \tau^\mp)$ under the assumption that
$\big| C^{L/R}_{\tau e} \big| =\big| C^{L/R}_{e \tau }\big| $.
The results in  Table \ref{table:Z:significance} yield the $2\sg$ upper bounds on $\br(Z \to e^\pm \tau^\mp) $ as
\bea
\left. \br(Z \to e^\pm \tau^\mp) \right|_{\rm LHeC} &<& 
	\left\{
	\begin{array}{ll}
	2.22\times 10^{-7} & \hbox{ for } \lumtot = 1\iab;\\
	1.49 \times 10^{-7} &  \hbox{ for } \lumtot = 3\iab;
	\end{array}
	\right.
\\[7pt] \nn
\left. \br(Z \to e^\pm \tau^\mp) \right|_{\rm FCC-he} &<& 
\left\{
	\begin{array}{ll}
	2.27 \times 10^{-7} & \hbox{ for } \lumtot = 1 \iab; \\
	9.91 \times 10^{-8} & \hbox{ for } \lumtot = 10 \iab.
	\end{array}
	\right.
\eea

The results at the LHeC are indeed impressive, 
especially when compared to the prospects at the HL-LHC. 
The recent bound set by the ATLAS Collaboration~\cite{ATLAS:2021bdj} 
with a total luminosity of $139\ifb$ is $\br(Z\to e\tau) < 5.0\times 10^{-6}$. 
If we extrapolate this bound to the HL-LHC with $\lumtot=3\iab$ but no background uncertainty, 
the expected limit becomes $\left. \br(Z\to e\tau) \right|_{\rm HL-LHC} \lsim 1.08 \times 10^{-6}$.
Notably, the LHeC 
exhibits an improvement by a factor of about five with respect to the HL-LHC.
Considering that both the HL-LHC and LHeC are expected to complete their programs around the same time scale, it is crucial not to overlook the potential of the LHeC in probing LFV in $Z$ boson decays.

In contrast, the bounds obtained at the FCC-he do not experience significant improvement, 
despite the substantial increase in proton beam energy.
This suggests that alternative strategies should be explored to effectively probe LFV in $Z$ boson decays at the FCC-he.
One promising approach is to raise the electron beam energy into $120\gev$~\cite{Jueid:2021qfs}. 
Therefore, serious consideration should be given to the option of a higher electron beam energy,
especially if our primary goal at the FCC-he is to achieve a more comprehensive exploration of BSM signals.

Finally, we discuss the limitations and potential of the \emph{indirect} probe of the LFV $Z$ decays.
Although we may observe the BSM signal of $p e^- \to \jf \tau^-$ in the future,
it does not provide conclusive evidence for LFV $Z$ decays. 
The observed signal could be explained by the presence of other new particles decaying into $e\tau$
such as $Z'$~\cite{Langacker:2008yv}, 
scalar neutrinos in the $R$-parity-violating supersymmetry models~\cite{Farrar:1978xj,Barbier:2004ez},
and quantum black holes in low scale gravity~\cite{Gingrich:2009hj}.
While disentangling the indirect signals of the LFV $Z$ from those of other BSM effects is beyond the scope of this paper,
we provide a qualitative discussion focusing on the discrimination between LFV $Z$ and $Z'$ bosons.

The recent heavy resonance search for a LFV $Z'$ boson by the CMS Collaboration~\cite{CMS:2022fsw}
excluded $M_{Z'} \lsim 4\tev$ 
in a benchmark scenario where the $Z'$ boson has identical couplings
to SM particles as the SM $Z$ boson but it can decay into $e\tau$ 
with an assumed branching ratio of $\br(Z'\to e\tau)=0.1$~\cite{CMS:2018hnz,ATLAS:2018mrn}.
This exclusion bound is in line with the constraints obtained from resonant $Z'$ searches in flavor-conserving dilepton final states, which also imply $M_{Z'} \gsim 4\tev$~\cite{ATLAS:2019erb}.
Therefore, let us consider the specific case of $M_{Z'} = 4\tev$ in our discussion.
If we assume a branching ratio of $\br(Z'\to e\tau)=2 \times 10^{-4}$,
the parton-level cross section of $ p e^- \to \jf \tau^-$ mediated by the $t$-channel $Z'$
is almost the same as that mediated by the LFV $Z$ boson with $C_{\tau e}^{L/R} = 10^{-3}$.
Consequently,  the signal rate alone cannot differentiate the indirect LFV signal of the SM $Z$ boson from that of $Z'$.

\begin{figure}[!t]
    \centering
    \includegraphics[width=\textwidth]{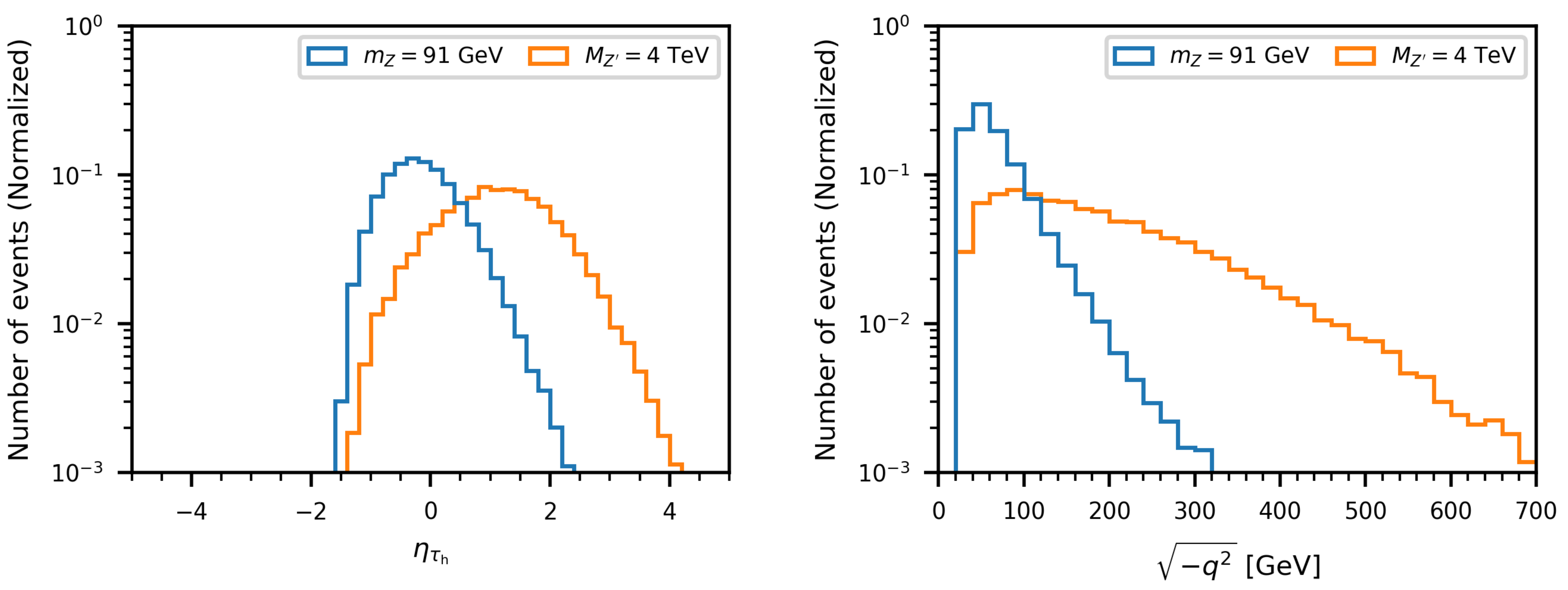}
    \caption{Normalized distributions against $\eta_\tauh$ (left panel) and $Q$ (right panel)
    for the process of $p e^- \to \jf \tau^-$ at the LHeC with $P_e=0$
    mediated by $Z$ (blue line) and $Z'$ with $M_{Z'}=4\tev$ (orange line).
The results are based on the dataset after the basic selection described in the text.
}
    \label{fig-ZZp}
    \end{figure}

However,
the significant mass difference between $m_Z$ and $M_{Z'}$ leads to distinct kinematic distributions.
In \fig{fig-ZZp},
we  highlight two distributions that exhibit pronounced differences, about $\eta_\tauh$ (left panel) and $\sqrt{-q^2}$ (right panel)
    for the process of $p e^- \to \jf \tau^-$ at the LHeC with $P_e=0$.
The distributions mediated by the LFV $Z$ and $Z'$ are depicted by the blue and orange solid lines, respectively.
Examining the $\eta_\tauh$ distribution, we observe that the outgoing $\tau$ from the LFV $Z'$ does not 
exhibit a preference for the backward direction.
This is attributed to the heavy mass of $Z'$,
which results in an effective four-point interaction involving $e$-$\tau$-$q$-$\bar{q}$.
As a result, the kinematic characteristics of the $t$-channel diagram are lost.
In addition,
significant discrepancies between the LFV $Z$ and $Z'$ processes are observed in the distributions of the momentum transfer, $\sqrt{-q^2}$.
In the case of the LFV $Z$ , we observe a sharper peak at a lower $\sqrt{-q^2}$ position
compared with the LFV $Z'$.
In summary, the distinctive kinematic features between the LFV $Z$ and $Z'$ provide optimism 
that it may be possible to disentangle these indirect signals when analyzing real data sets.

\subsection{Results of the LFV Higgs boson}
\label{sec:analysis:Z}
%%%%%%%%%%%%%%%%%%%%%%%%%%%%%%%%%%%%%%%%%%%%%%%%%%%%%

Exploring the LFV of the Higgs boson at electron-proton colliders poses challenges 
primarily due to the inherently low cross section of the Higgs production.
As summarized in Table \ref{table:Xsec:SM}, even  at the parton level,
the highest possible cross section via the CC production with $P_e=-80\%$
is merely about $145\fb$ ($604\fb$) at the LHeC (FCC-he).
Considering the current bounds of $\br(H\to e\tau/\mu\tau) \lsim 10^{-3}$,
the initial dataset with $\lumtot=1\iab$ consists of only a few hundred events, leading to a limited statistical sample.
Furthermore, the primary background arising from the CC production of the $Z$ boson, 
followed by $Z \to \tau_e \tau_h$, shares the same Feynman diagram topology  as the LFV Higgs signal. 
This similarity further aggravates the situation in background reduction. 
Despite our concerted efforts in a cut-based analysis to enhance the signal significance, we were only able to achieve a modest significance of $\mcs \simeq 1$.
 
To address these limitations, we turned to a multivariate analysis approach, 
utilizing the BDT algorithm~\cite{Roe:2004na} implemented through the \textsc{XGBoost} package~\cite{Chen:2016btl}.
This technique allows us to exploit the correlations among multiple observables 
and to maximize the discrimination power between signal and background events. 
The reference signal points are defined by
\bea
\br(H\to \ell^\pm \tau^\mp) = 10^{-3}, 
\eea
where $\ell^\pm=e^\pm,\mu^\pm$.

Let us first discuss the LFV decay of the Higgs boson in the $e\tau$ mode.
As discussed in the previous section,
we concentrate on the process
\bea
p e^- \to H + \jf \nu \to e^+ \tau^- + \jf \nu.
\eea
For the basic selection, we apply the following minimal conditions to avoid excessively suppressing the signal events:
\bit
\item We require $N_\tauh \geq 1$, $N_e \geq 1$, and $N_j \geq 1$,
where $N_X$ is the number of the $X$ object with $p_T>20\gev$, $|\eta_{j,\tauh}|<4.5$, and $|\eta_e|<3.5$.
\item We demand that the charge of the leading electron object be positive.
\eit
After the basic selection, the background from $j_e j_\tauh +\jf\nu$ can be neglected
due to the tiny mistagging probabilities of $P_{j\to e}$ and $P_{j\to \tauh}$.

To discriminate the signal from the backgrounds,
we use the following 19 observables as the BDT inputs:\footnote{In the literature, it is common practice to exclude highly correlated kinematic variables in BDT analyses. 
Although we observed that three of the parameters in our analysis exhibit correlations higher than 80\% in 
the signal, we decided to include all 19 variables.
Surprisingly, this choice resulted in a slight improvement of about $\mco(1)\%$ in the BDT performance. } 
\bea
\label{eq:BDT:H:variables}
&&N_j, \quad N_{\tauh}, \quad N_{e},\quad p_T^j,\quad p_T^\tauh,\quad p_T^e,
\quad \eta_j,\quad \eta_\tauh,\quad \eta_e ,\\ \nn
 &&\Dt R(\tauh,e) ,\quad \Dt R(j, \langle \tauh e \rangle ) ,\quad 
\Dt\phi(\tauh,e) ,\quad M_{\tauh e},\quad p_T^{\langle \tauh e \rangle}, \quad \met,
\\ \nn
&&\Dt \phi( \vec{p}_{\rm miss},\tauh),\quad \Dt \phi( \vec{p}_{\rm miss},e),\quad \Dt \phi( \vec{p}_{\rm miss},\langle \tauh e \rangle),\quad 
m_{\rm col}.
\eea
Here  $\langle \tauh e \rangle$ denotes the system consisting of $\tauh^-$ and $e^+$,
of which the momentum is the vector sum of $\vec{p}_\tauh$ and $\vec{p}_e$.
$M_{\tauh e}$ is the invariant mass of the $\langle \tauh e \rangle$ system, and 
$\vec{p}_{\rm miss}$ is the negative vector sum of the momenta of all the observed particles.
$m_{\rm col}$ is the collinear mass of the $\langle \tauh e \rangle$ system, given by
\bea
m_{\rm col} = \frac{M_{\tauh e}}{\sqrt{x_\tauh^{\rm vis}}},
\eea
where $x_\tauh^{\rm vis} $ is  defined by
\bea
\frac{1}{x_\tauh^{\rm vis} } =1+\frac{\vec{p}_{\rm miss} \cdot \vec{p}_\tauh}{\lf p_T^\tauh \ri^2}.
\eea
The area under the receiver operating characteristic curve is used as the metric to evaluate the BDT performance.
The end of the training is set by \texttt{early\_stopping\_rounds=5} in the \textsc{XGBoost}.

For $H\to\mu\tau$ mode,
we consider the CC process of $H\to \mu^\pm \tau^\mp$.
The final state consists of one muon, one  tau lepton, one forward jet, and missing transverse energy.
We do not impose any condition on the electric charge of the muon.
The basic selections require $N_\tauh \geq 1$, $N_\mu \geq 1$, and $N_j \geq 1$
where the object satisfies $p_T>20\gev$, $|\eta_{j,\tauh}|<4.5$, and $|\eta_\mu|<3.5$.
As BDT inputs, we used the same kinematic observables in \eq{eq:BDT:H:variables} with the positron replaced by the muon.

Let us outline our simulation procedures for the LHeC first.
For $H\to e^+\tau^-$ mode, we generated $2.0\times 10^6$ events for the signal and $2.0\times 10^7$ events for each background. 
The training dataset consisted of $3.3\times 10^5$ events for the signal and $3.5\times 10^5$ events for the background.
Similarly, for $H \to \mu^\pm\tau^\mp$, 
we generated $2.0\times 10^6$ signal events and $2.0\times 10^7$ background events, 
followed by training with $6.7\times 10^5$ signal events and $7.2\times 10^5$ background events. 
The dataset was divided into 60\% for training, 20\% for testing, and 20\% for validating the algorithm.
The validation has been successfully accomplished.

\begin{figure}[!h]
    \centering
    \includegraphics[width=0.9\linewidth]{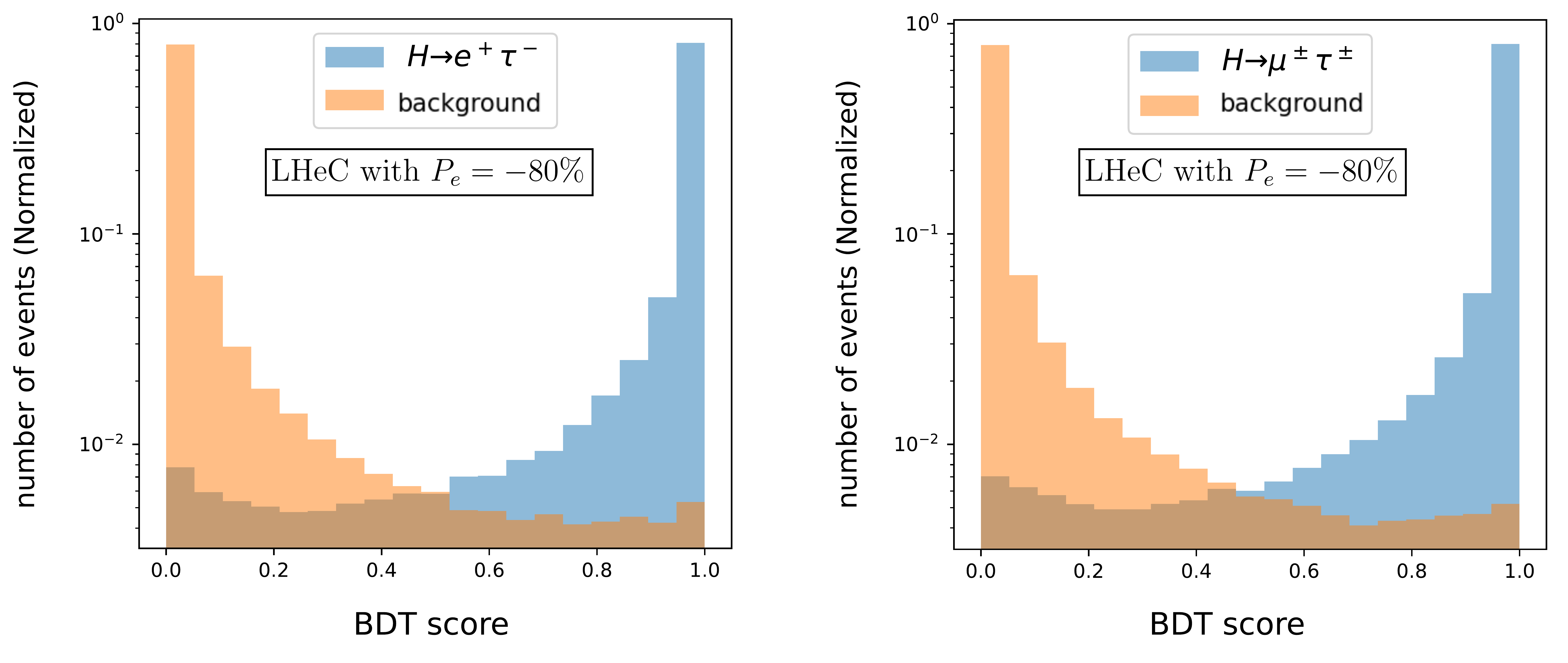}\\
    \caption{Normalized signal and background distributions 
    against BDT scores. The left panel is for $H\to e^+ \tau^-$ 
    and the right panel is for $H \to \mu^\pm\tau^\mp$ with $\br(H\to e^\pm \tau^\mp/\mu^\pm \tau^\mp) = 10^{-3}$
     at the LHeC with  $P_e =-80\%$.
    The results are based solely on the testing dataset after the basic selection. 
}
    \label{fig-H-BDT}
    \end{figure}

To demonstrate the performance of our approach, we present
in \fig{fig-H-BDT} the normalized distributions of the signal and backgrounds against the BDT response at the LHeC with $P_e =-80\%$.
The results are based solely on the testing dataset after the basic selection. 
Our dataset is sufficiently large, ensuring that the distribution from test samples closely matches that from training samples.
The left panel represents the results for $H\to e^+ \tau^-$, while the right panel displays the results for $H \to \mu^\pm\tau^\mp$. 
The plot clearly shows a distinct separation between the signal and background distributions, 
indicating the effectiveness of our approach in discriminating LFV signals..
We followed a similar approach for the FCC-he, 
where the results demonstrate a comparable behavior. 
 
 \begin{figure}[!h]
    \centering
    \includegraphics[width=0.6\linewidth]{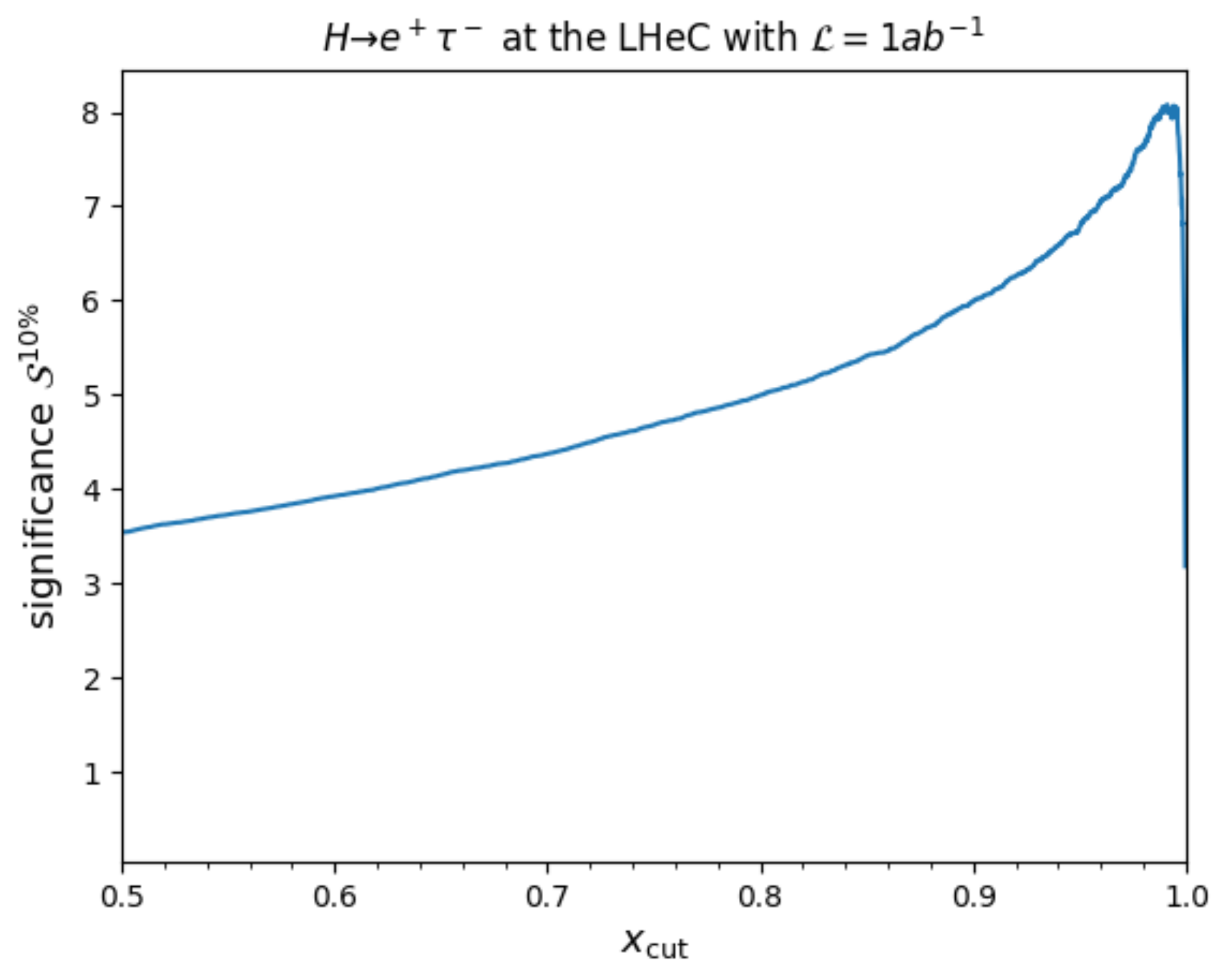}\\
    \caption{Significance of the signal $H \to e^+ \tau^-$
    against the BDT score cut $\xc$ at the LHeC with $P_e=-80\%$ and the total luminosity of $1\iab$. 
We include a 10\% background uncertainty.}
    \label{fig-significancen-H-xcut}
    \end{figure}

Given the rarity of our signal, 
it is crucial to carefully select the BDT score cut $\xc$. 
In \fig{fig-significancen-H-xcut}, 
we illustrate the signal significance as a function of $\xc$ for the $H \to e^+ \tau^-$ process at the LHeC with $\lumtot=1\iab$ and $P_e = -80\%$. 
We observe that as $\xc$ increases,  the significance increases, reaching a maximum before eventually decreasing.
We select $\xc$ to maximize the significance while ensuring that at least one background event remains.

Brief comments are warranted on the decreasing behavior of the significance as $\xc\to 1$,
 which corresponds to the limit  $\nbg \to 0$.
In \eq{eq:significance:nbg}, 
the significance expression exhibits divergence as $\nbg$ approaches 0,
which might seem inconsistent with the findings shown in \fig{fig-significancen-H-xcut}.
However, this divergence is a consequence of the incorrect assumption that
$\nsg$ remains fixed even in the limit of $\nbg \to 0$,
which is not reflective of real-world high-energy collider experiments.
In practice, the background events can be almost entirely eliminated
only when we impose very stringent cuts. 
However, this stringent filtering also leads to a comparable reduction in the signal events. 
As a result, the significance decreases rather than diverging.

To illustrate this point, let us consider an equivalent scenario of $\nbg \ll 1$ by decreasing the luminosity. 
Suppose we start with a given luminosity $\lumtot^0$ and have $\nbg=1$ and $\nsg=1000$, resulting in a significance of 108.  
Now, if we reduce the luminosity to $0.01 \lumtot^0$ so that
$\nbg$ takes on a very small value,
both  $\nsg$ and $\nbg$ decrease.
As a result, the significance is reduced to 10.8, clearly illustrating a decrease rather than an increase.

\begin{table}
%\begin{ruledtabular}
  {\renewcommand{\arraystretch}{1.2}
  \begin{tabular}{|c|c|c|c|c|c||c|c|c|c|c|}
\toprule
\multicolumn{11}{|c|}{$\br(H\to e\tau/\mu\tau)=10^{-3}$}\\ \toprule
 & \multicolumn{5}{c||}{LHeC with $P_e=-80\%$} & \multicolumn{5}{c|}{FCC-he with $P_e=-80\%$} \\ \hline
 & $\lumtot$ & $\xc$ & $\nsg$ & $\nbg$  & $\mcs^{10\%}$ &  $\lumtot$ & $\xc$ & $\nsg$ & $\nbg$  & $\mcs^{10\%} $ \\ \hline
\multirow{4}{*}{~~$H\to e^+ \tau^-$~~} & ~~\multirow{2}{*}{$1\iab$}~~ & ~0.978~ & ~15.4~ & ~1.0~  & ~7.61~  & ~~\multirow{2}{*}{$1\iab$}~~ & ~0.993~ &  ~47.3~ & ~1.4~   & ~~15.23~~ \\ \cline{3-6}\cline{8-11}
& & 0.900 & 18.1 & 4.1 & 5.99 &   & 0.900 & 77.2 & 22.1 & 10.15 \\ \cline{2-11}
& \multirow{2}{*}{$3\iab$} & 0.991 & 38.9 & 1.3 & ~13.64~ &  \multirow{2}{*}{$10\iab$}  &  0.993 & ~472.5~ & ~13.9~  & ~~39.19~~\\ \cline{3-6}\cline{8-11}
&  & 0.900 & 54.3 & 12.3 & ~9.68~ &  &  0.900 & 772.1 & 221.2  & 17.47  \\ \hline
\multirow{4}{*}{$H\to \mu^\pm \tau^\mp$} & \multirow{2}{*}{$1\iab$}  & 0.990 & 26.5 & 1.0 & 11.01   & \multirow{2}{*}{$1\iab$} & 0.992 & ~103.2~ & ~3.9~ & 20.62  \\ \cline{3-6}\cline{8-11}
 &  & 0.900 & 36.3 & 8.7 & 8.02  &  & 0.900 & 156.1 & 45.9 & 12.61  \\ \cline{2-11}
& \multirow{2}{*}{$3\iab$}  & 0.994 & 69.3 & 1.9 & 18.43 & \multirow{2}{*}{$10\iab$} & 0.997 & 713.4 & 13.9 & 49.75  \\ \cline{3-6}\cline{8-11}
& & 0.900 & 109.0 & 26.1 & 12.29     &  & 0.900 & 1561 & 459.4 & 18.28   \\ \bottomrule
\end{tabular}
}
\caption{\label{table:H:significance}
The BDT analysis results for the signals of $H\to e^+ \tau^-$ and $H\to \mu^\pm \tau^\mp$
at the LHeC and FCC-he with $P_e=-80\%$, including 10\% background uncertainty.
Here $\nsg$ ($\nbg$) is the number of signal (background) events and
$\xc$ is the cut on the BDT output of the testing dataset
}
\end{table}

In Table \ref{table:H:significance},
we present the results of the BDT analysis for the LFV $H$ decays,
including the values of $\xc$, $\nsg$, $\nbg$,  and
the significances for the signals of $H\to e^+ \tau^-$ and $H\to \mu^\pm \tau^\mp$ 
with $\br(H\to e\tau/\mu\tau)=10^{-3}$
at the LHeC and FCC-he.
We provide two sets of results: one based on the optimal $\xc$ obtained by maximizing the significance while ensuring at least one background event, and the other using a reference value of $\xc=0.9$.
In our analysis, we account for a 10\% background uncertainty. 
We also consider two cases of the total integrated luminosity: $1\iab$ and $3\iab$ for the LHeC, and $1\iab$ and $10\iab$ for the FCC-he.

Our optimal choice of $\xc$ yields remarkably high signal significances for the LFV decay modes $H\to e\tau$ and $H\to\mu\tau$, making them sufficiently strong to claim a discovery if $\br(H\to e\tau/\mu\tau)=10^{-3}$.
The LHeC with the proposed total luminosity of $1\iab$
yields $\mcs^{10\%} \simeq 7.6$ for $H\to e^+\tau^-$ and $\mcs^{10\%} \simeq 11.0$ for $H\to \mu^\pm\tau^\mp$.
Tripling the total luminosity into $\lumtot=3\iab$
enhances the significance proportional to $\sqrt{\lumtot}$ even with the 10\% background uncertainty.
It is attributed to the very small background events ($\nbg \simeq 1$).
At the FCC-he with the total luminosity of $1\iab$,
the significances are much higher:
$\mcs^{10\%}_{1\iab} \simeq 15$ for $H\to e^+\tau^-$ and $\mcs^{10\%}_{1\iab} \simeq 21$ for $H\to \mu^\pm\tau^\mp$.
The improvement from the high luminosity of $10\iab$ is impressive.
We can accommodate 
$\mcs^{10\%}_{1\iab} \simeq 39$ for $H\to e^+\tau^-$ and $\mcs^{10\%}_{1\iab} \simeq 50$ for $H\to \mu^\pm\tau^\mp$.
Notably, the results for a reference value of $\xc=0.9$ do not show a substantial decrease in the signal significance, 
illustrating the efficiency of our analysis. 
This indicates that our approach remains effective even for slightly suboptimal choices of the $\xc$ parameter.

Based on the results in Table \ref{table:H:significance},
we calculate the $2\sg$ upper bounds on the LFV decays of the Higgs boson as
\bea
\br(H\to e^\pm \tau^\mp) &<& \left\{
{\renewcommand{\arraystretch}{1.2} 
 \begin{array}{ll}
		1.7 \; (0.74)\times 10^{-4}, & \hbox{ at the LHeC with }~ \lumtot = 1\;(3)\iab,\\
		6.3 \; (1.9)\times 10^{-5}, & \hbox{ at the FCC-he with }~ \lumtot = 1\; (10)\iab; \\
		\end{array}
		}
		\right. \\ \nn
\br(H\to \mu^\pm \tau^\mp) &<& \left\{
{\renewcommand{\arraystretch}{1.2} 
 \begin{array}{ll}
		1.0\;(0.49)\times 10^{-4}, & \hbox{ at the LHeC with }~ \lumtot = 1\;(3)\iab,\\
		4.5\;(1.2) \times 10^{-5}, & \hbox{ at the FCC-he with }~ \lumtot = 1\; (10)\iab. \\
		\end{array}
		}		
		\right.
\eea
It is noteworthy that the LHeC with a total luminosity of $1\iab$ 
can establish higher sensitivities to the LFV decay branching ratios of the Higgs boson 
than the HL-LHC with $\lumtot=3\iab$. 
Furthermore, the performance of the FCC-he is particularly remarkable, 
as the dataset with $\lumtot=1\iab$ has the capacity to probe LFV decay branching ratios of the Higgs boson 
as low as $\mco(10^{-5})$.

  \section{Conclusions}
 \label{sec:Conclusions}
 
  We have conducted a comprehensive investigation into the potential of  the LHeC and FCC-he for detecting lepton flavor violation (LFV) phenomena
 of the $Z$ and Higgs bosons. 
 For our analysis, we considered collision energies of $E_e =50\gev$ and $E_p=7\tev$ for the LHeC 
 and $E_e =60\gev$ and $E_p=50\tev$ for the FCC-he. 
 We found that electron-proton colliders are well-suited for detecting LFV phenomena 
thanks to negligible pileups, small QCD backgrounds, and the feasibility to distinguish the charged-current
from neutral current processes.

For the $Z$ LFV study, 
we focused on the indirect probe $p e^- \to \jf \tau^-$ mediated by the $Z$ boson in the $t$-channel. 
%Direct searches for on-shell decay of $Z \to e\tau/\mu\tau$ face substantial challenges 
%due to their small branching ratios below about $10^{-6}$.  
Employing a conventional cut-based analysis and consider{ing a full detector simulation with 10\% background uncertainty, 
 we have demonstrated that the LHeC with $\lumtot=1\iab$ and electron beam polarization of $P_e=0$ 
 can yield a significance of about 7.08 if $C_{\tau e}^L =C_{\tau e}^R=10^{-3}$, while the FCC-he with $\lumtot=1\iab$ has a significance of about 8.04.
We have extended the analysis employing a dedicated multivariate analysis utilizing the BDT algorithm with the \textsc{XGBoost} package,
which enhances the significances by about 74\% (54\%) at the LHeC (FCC-he) with $\lumtot=1\iab$.
 We obtained $2\sg$ bounds of $\br(Z\to e\tau)\leq 2.22~(2.27) \times 10^{-7}$ at the LHeC (FCC-he) with $\lumtot=1\iab$.
Remarkably, the LHeC with a luminosity of $\lumtot = 1\iab$ alone exhibits higher sensitivity compared to the HL-LHC with $\lumtot = 3\iab$.  

For the LFV studies involving the Higgs boson, 
we have concentrated on the direct observation of on-shell decays. 
Specifically, we have emphasized  $H\to e^+\tau^-$ to avoid the electron-related backgrounds. 
To address the challenges associated with the small production cross sections of the Higgs boson at the LHeC and FCC-he, 
we employed a multivariate analysis utilizing the BDT algorithm with the \textsc{XGBoost} package.
Our analysis yielded the $2\sg$ bounds  at the LHeC (FCC-he) with $\lumtot=1\iab$ and $P_e=-80\%$
such that  $\br(H\to e^\pm \tau^\mp) <1.7 \times 10^{-4} \lf6.3 \times 10^{-5}\ri$ and $\br(H\to \mu^\pm \tau^\mp) < 1.0 \times 10^{-4} \lf 4.5 \times 10^{-5} \ri$.
These bounds surpass the projected sensitivities of the HL-LHC with $\lumtot = 3\iab$.

Our investigation into the LFV signatures of the $Z$ and Higgs bosons 
has unveiled the high potential of the LHeC and FCC-he in observing these rare processes. 
Given the impressive outcomes of our study, we strongly endorse and support the LHeC and FCC-he programs.

\section*{Acknowledgments}
We express our gratitude to the referee for the invaluable comments and thought-provoking questions, 
which significantly enhanced the quality and impact of our paper.
This paper was written as part of Konkuk University's research support program for its faculty on sabbatical leave in 2023.
The work of AJ is supported by the Institute for Basic Science (IBS) under the project code, IBS-R018-D1.
The work of JK, SL, JS, and DW is supported by 
the National Research Foundation of Korea, Grant No.~NRF-2022R1A2C1007583.

%\bibliographystyle{JHEP}
%\bibliography{./3PRD-LFV-Z-H-ep}

\end{document}